\documentclass[12pt]{article}
\pdfoutput=1

\usepackage{putex}
\usepackage{graphicx}
\usepackage{caption}
\usepackage{amsmath}
\usepackage{amssymb}
\usepackage{array}
\usepackage{bm}
\usepackage{multirow}
\usepackage{mathtools}
\usepackage{comment}
\usepackage{subcaption}
\usepackage{epstopdf}
\usepackage{enumerate}
\usepackage{cite}
\usepackage{youngtab}
\usepackage{tensor}
\usepackage{slashed}
\usepackage[aligntableaux=center]{ytableau}
\usepackage[utf8]{inputenc}
\usepackage{rotating}
\usepackage{bigfoot}
\usepackage[
      colorlinks=true,
      linkcolor=blue,
      urlcolor=blue,
      filecolor=black,
      citecolor=red,
      linktocpage=true
      ]{hyperref}
\usepackage{dsfont}

\newcommand {\be} {\begin {equation}}
\newcommand {\ee} {\end {equation}}
\newcommand {\nn} {\nonumber}

\newcommand {\bes} {\begin {equation*}}
\newcommand {\ees} {\end {equation*}}

\newcommand{\es}[2] {\begin{equation} \label{#1} \begin{split} #2 \end{split} \end{equation}}

\newcommand{\R}{\mathbb{R}}

\newcommand{\cA}{{\mathcal A}}
\newcommand{\cB}{{\mathcal B}}
\newcommand{\cC}{{\mathcal C}}

\newcommand{\cG}{{\mathcal G}}

\newcommand{\cN}{{\mathcal N}}
\newcommand{\cO}{{\mathcal O}}

\newcommand{\rmi}{i}
\newcommand{\rme}{\mathrm{e}}
\newcommand{\rmd}{\mathrm{d}}

\newcommand{\bea}{\begin{equation}\begin{aligned}}
\newcommand{\eea}[1]{\label{#1}\end{aligned}\end{equation}}

\newcommand{\beq}{\begin{equation}}
\newcommand{\eeq}{\end{equation}}

\def\ie{\begin{equation}\begin{aligned}}
\def\fe{\end{aligned}\end{equation}}

\numberwithin{equation}{section}

\def\<{\langle}
\def\>{\rangle}

\usepackage{tikz}
\usetikzlibrary{decorations.pathmorphing, decorations.markings}
\usepackage{subcaption}

\begin{document}

\preprint{}

\institution{imperial}{Abdus Salam Centre for Theoretical Physics, Imperial College London, London SW7 2AZ, UK}

\title{Extremal couplings, graviton exchange, and gluon scattering in AdS}

\authors{Shai M.~Chester,\worksat{\imperial} Rishi Mouland,\worksat{\imperial} and Jesse van Muiden\worksat{\imperial}  }

\abstract{
Extremal cubic couplings in AdS relate bulk fields such that $\Delta_i+\Delta_j=\Delta_k$. Such couplings lead to divergent 3-point Witten diagrams, and do not occur in theories with maximal supersymmetry. We consider the simplest theories where such coupling are non-zero, which is type IIB string theory with $N$ D3 branes probing various configurations of sevenbranes, which are dual to certain 4d $\mathcal{N}=2$ SCFTs. At large $N$, the low energy effective theory is supergravity on $AdS_5\times S^5$ with a singularity, whose fixed point locus is $AdS_5\times S^3$. These theories have infinite towers of graviton modes, as well as gluon modes on the sevenbranes. We compute the nonzero coupling between these modes, which is in general (super)-extremal. We use these couplings to compute the graviton exchange term in the holographic correlator of gluon KK modes $\langle22pp\rangle$, which appears at the same order $1/N^2$ as 1-loop gluon exchange, and receives contributions from a whole tower of graviton modes. We use this graviton exchange term to compute the unmixing of the single trace graviton modes with double traces of gluon modes, which explains the divergent 3-point diagrams. Finally, for $\langle 2222\rangle$ for the simplest 4d $\mathcal{N}=2$ gauge theory, we use supersymmetric localization and the new graviton exchange term to completely fix  the correlator at order $1/N^2$.
}
\date{}

\maketitle

\tableofcontents
\newpage

\section{Introduction}\label{sec:intro}

\begin{figure}[h]\centering
	\begin{tikzpicture}[scale=0.8]

	\draw[dashed] (0,0) circle (3);

	\coordinate (bulk) at (0,0);

	\filldraw (bulk) circle (2pt)  node[below right] {$\beta_{ijk}$};

	\coordinate (A) at (30:3);  
	\coordinate (B) at (150:3); 
	\coordinate (C) at (270:3); 

	\draw[thick] (bulk) -- (A);
	\draw[thick] (bulk) -- (B);
	\draw[thick] (bulk) -- (C);

	\node at (30:3.4) {$x_k$};
	\node at (151:3.4) {$x_i$};
	\node at (273:3.4) {$x_j$};

	\node at (138:1.7) {$\phi_i$};
	\node at (18:1.7) {$\phi_k$};
	\node at (258:1.7) {$\phi_j$};

	\end{tikzpicture}
	\caption{Witten diagrams for the cubic interactions between three scalars.}\label{Fig: Witten diagrams for the cubic couplings}
\end{figure}
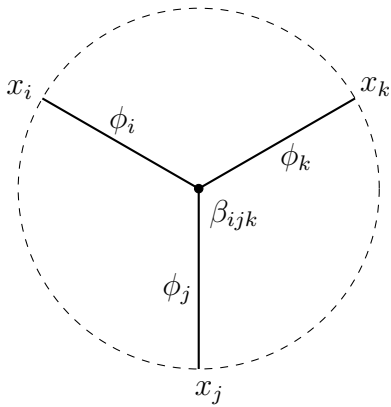

The AdS/CFT dictionary relates correlation functions in a $d$-dimensional conformal field theory (CFT) to Witten diagrams in $d+1$-dimensional Anti-de Sitter space (AdS$_{d+1}$) \cite{Maldacena:1997re}. The simplest example of this correspondence relates the coefficient $\lambda_{ijk}$ of a three point function (i.e. the OPE coefficient) of scalar operators with scaling dimensions $\Delta_i$ in the CFT to the corresponding bulk coupling $\beta_{ijk}$ in AdS$_{d+1}$ as  \cite{Witten:1998qj,Freedman:1998tz}
\es{3point}{
\lambda_{ijk}=\beta_{ijk}\frac{\pi ^{d/2}\sqrt{\cC_{\Delta_1}\cC_{\Delta_2}\cC_{\Delta_3}}  \Gamma(\frac{\Delta_i+\Delta_j-\Delta_k}{2})  \Gamma(\frac{\Delta_k+\Delta_i-\Delta_j}{2})  \Gamma(\frac{\Delta_j+\Delta_k-\Delta_i}{2})
 \Gamma(\frac{\Delta_i+\Delta_j+\Delta_k-d}{2})
}{2 \Gamma(\Delta_i) \Gamma(\Delta_j) \Gamma(\Delta_k)}\,,
}
where $\cC_\Delta=\frac{\Gamma(\Delta)}{2\pi^{d/2}\Gamma(\Delta+1-d/2)}$ and the coefficient comes from the vertex integral depicted in Figure \ref{Fig: Witten diagrams for the cubic couplings}. This coefficient diverges whenever $\Delta_i=\Delta_j+\Delta_k+2a$ for $a=0,1,2,\dots$, in which case we call the coupling extremal (for $a=0$) or super-extremal (for $a>0$). 
This divergence is surprising, as all examples of AdS/CFT include towers of KK modes dual to single trace operators with protected integer $\Delta_i$, which arise from compactifying string theory down to AdS$_{d+1}$, that have nonzero finite three-point functions that are extremal. For instance, consider the duality between $N$ D3 branes in IIB string theory and 4d $\mathcal{N}=4$ $SU(N)$ super-Yang-Mills (SYM), where at large $N$ the bulk is described by supergravity on $AdS_5\times S^5$. Compactifying the 10d graviton on $AdS_5\times S^5$ gives KK modes dual to half-BPS operators in the CFT with $\Delta=2,3,4\dots$, and $\lambda_{ij (i+j)}$ are kinematically allowed  \cite{Lee:1998bxa,DHoker:1998ecp,Liu:1999kg,DHoker:1999jke}. The resolution in this case is that $\beta_{ij(i+j)}$ are all zero, and the bulk fields are dual to linear combinations of single and higher trace operators whose $\lambda_{ij (i+j)}$ also vanish \cite{Aprile:2020uxk,Aprile:2018efk}. The vanishing bulk couplings explain why the four-point function of the $\Delta=2$ graviton multiplet at large $N$ only receives contributions from the exchange of the graviton multiplet itself \cite{Arutyunov:2000py}, even though the extremal $\Delta=4$ KK mode would have been allowed by symmetry.

Extremal bulk couplings in fact vanish in all examples of AdS/CFT with maximal supersymmetry.\footnote{They were also recently shown to vanish in a case with less than maximal supersymmetry \cite{Bobev:2025gzu}.} This might give the mistaken impression that these bulk couplings always vanish. A recent paper \cite{Castro:2024cmf} showed from the bottom up perspective how extremal bulk couplings need not vanish, and that the divergence in \eqref{3point} should instead be regularized and interpreted as mixing between the single trace operator of dimension $\Delta_i = \Delta_j + \Delta_k+2a $ and a particular double trace operator of the same dimension built from the operators of dimension $\Delta_j$ and $\Delta_k$. In fact, extremal couplings are ubiquitous in top-down holography with less than maximal supersymmetry. In this paper we will discuss the simplest examples with 4d $\mathcal{N}=2$ supersymmetry, which involves $N$ D3 branes and a finite number of sevenbranes. We will show that an infinite set of extremal and super-extremal bulk couplings between graviton modes and gluons on the sevenbranes are nonzero, and that they contribute an infinite set of exchange diagrams to scattering of gluons at large $N$.

\begin{table}[h]
    \centering
    \renewcommand{\arraystretch}{1.5} 
    \begin{tabular}{c | c c c c c c}
    	$ G_F $ & $A_1$ & $A_2$ & $D_4$ & $E_6$ & $E_7$ & $E_8$ \\\hline
    	$\tau$ & $i$ & $e^{\pi i /3}$ & - & $e^{\pi i /3}$ & $i$ & $e^{\pi i /3}$\\
        ${\bm\Delta}$ & $4/3$ & $3/2$ & $2$ & $3$ & $4$ & $6$ 
    \end{tabular}
    \caption{ADE classification of D3-branes probing F-theory singularities.}\label{Tab: D7 brane singularities}
	\end{table}
\noindent 
In more detail, we consider the class of F-theory compactifications where the complexified string coupling $\tau$ takes a constant value \cite{Sen:1996vd,Dasgupta:1996ij}. These arise from $N$ D3 branes probing F-theory singularities labeled by group $G_F$, which arise from $N_7$ sevenbranes, and are summarized in Table \ref{Tab: D7 brane singularities}.\footnote{There is another theory with constant $\tau$ that has an $A_0$ type singularity, but we will not consider it because there is no gluon scattering in this theory.} At large $N$, the low energy effective theory is given by supergravity on $AdS_5\times S^5$, but with a singularity whose fixed point locus is super-Yang-Mills (SYM) on $AdS_5\times S^3$ with gauge group $G_F$. The holographic dual \cite{Fayyazuddin:1998fb,Aharony:1998xz} of these theories are given by 4d $\mathcal{N}=2$ CFTs with the flavor symmetry $G_F\times SU(2)_L$, R-symmetry $SU(2)_R\times U(1)_R$, and the lowest Coulomb branch operator with scaling dimension ${\bm\Delta} = \frac{1}{1-N_7/12}$ \cite{Aharony:2007dj}. Compactifying the D7 branes on $S^3$ gives an infinite tower of gluon modes dual to scalar single trace superprimaries with $\Delta=p$ for $p=2,3,\dots$ that transforms in the adjoint of $G_F$ and the $(\frac p2-1,\frac p2)_0$ of $SU(2)_L\times SU(2)_R\times U(1)_R$.\footnote{We denote irreps of $SU(2)_L\times SU(2)_R$ by their isospin.} Compactifying the 10d graviton gives several infinite towers of graviton modes. For instance, for each irrep $(\frac r2-1,\frac r2-1)_0$, there is a tower of KK modes dual to scalar single trace superprimaries with $\Delta=k_r$ for $k_r=r,r+2,r+4\dots$ that are singlets under $G_F$.\footnote{Other graviton KK mode tower are related to chiral operators charged under $U(1)_R$. These towers can be seen by decomposing half-BPS multiplets of $\mathcal{N}=4$ to $\mathcal{N}=2$ such that $SU(4)_R\to SU(2)_R\times SU(2)_L\times U(1)_R$.} For instance, $k_2=2$ corresponds to the 5d graviton itself that is dual to the protected stress tensor multiplet superprimary, while all other $k_r$ are dual to long multiplets. We compute all nonzero bulk couplings $\beta_{pqk_r}$ of two gluon modes and a graviton mode,\footnote{There are also half-BPS graviton modes transforming as $(\frac p2-1,\frac p2-1)_0$ with $\Delta=p-2$ for $p>3$, but these have zero cubic bulk coupling with $2$ and $p$, so we do not consider them. They do couple to $p$ and $p$ for $p>2$ \cite{Chester:2025ssu}, but this is not relevant to $\langle 22pp\rangle$ since they do not couple to $2$ and $2$.} where $r=|p-q|+2,\dots,p+q-2$, which is generically (super-)extremal.

We then consider scattering of two $\Delta=2$ and two $\Delta=p$ gluon modes on $AdS_5\times S^3$, which is dual to the correlator $\langle 22pp\rangle$ of the corresponding half-BPS multiplets. The analytic bootstrap \cite{Rastelli:2017udc} (i.e. crossing, analyticity, and flat space limit) restricts this correlator at large $N$ and finite $\tau$ to take the form 
\es{mellin_generalP}{
M=&\frac{1}{N}M_{F^2}+\frac{1}{N^2}\left[M_R+M_{F^2|F^2}(s,t)+\sum_i(b^i_{F^4,s} M^i_{F^4,s}+b^i_{F^4,tu} M^i_{F^4,tu})\right] \\
&+\frac{\log N}{N^2}(b_{\log,s}\,M_{\log,s}+b_{\log,tu}\,M_{\log,tu})+O(N^{-5/2})\ ,
}
where $i$ runs over the number of quartic Casimirs of the theory, which is three for the $D_4$ theory and one otherwise. The leading term $M_{F^2}$ corresponds to exchange of the gluon itself and was fixed in \cite{Alday:2021odx}. The 1-loop gluon exchange term $M_{F^2|F^2}$ was computed in \cite{Alday:2021ajh} using the AdS unitarity cut method \cite{Aharony:2016dwx}, while integrated constraints were used in \cite{Behan:2023fqq,Behan:2024vwg} to compute the logarithmic threshold contact term $b_{\log}\equiv b_{\log,s}=b_{\log,tu}$ for $p=2$ for all theories, as well as the higher derivative contact term $b_{F^4}\equiv b_{F^4,s}=b_{F^4,tu}$ for the $D_4$ theory. In this paper we consider the graviton exchange term $M_R$, which gets contributions from the exchange of graviton modes for all $k_2=2,4,6,\dots$ in the direct channel, and $k_p=p,p+2,p+4,\dots$ in the cross channel. Using the bulk couplings $\beta_{ppk_2}$ and $\beta_{2pk_p}$, we find the surprisingly simple answer
  \es{MR2}{
 \hspace{-.1in}  M_{R}&=-\frac{p}{(p-2)!\bm\Delta}\Bigg[\Big[(p+1)H_{1-\frac{s}{2}}+\frac{4}{s-2}\Big]\delta^{AB}\delta^{CD}+\Big[(p+1)H_{\frac{2+p-u}{2}}+\frac{2p}{u-p}\Big]\delta^{AC}\delta^{BD}\\
   &\qquad\qquad\qquad\qquad\qquad\qquad\qquad\qquad\qquad\quad+\Big[(p+1)H_{\frac{2+p-t}{2}}+\frac{2p}{t-p}\Big]\delta^{AD}\delta^{BC}\Bigg]\,,\\
   }
   where $H_x$ is a harmonic series, $A,B,C,D$ are $G_F$ adjoint indices, and the correlator is written in Mellin space, as will be explained in the main text.  We also explain how to unmix the single trace gravitons for $k_2\geq4$ and double trace operators formed from gluon KK modes,\footnote{This mixing should not be confused with the mixing discussed in \cite{Aprile:2020uxk,Aprile:2018efk}, which concerns the definition of bulk fields in terms of operators described in the free theory. For instance, the graviton mode in our case is unprotected and so its OPE coefficient at strong coupling cannot be computed from the free theory, unlike the OPE coefficients of the half-BPS operators considered in \cite{Aprile:2020uxk,Aprile:2018efk}.} so that we can compute the $1/N$ contribution to the anomalous dimensions of these operators in $\langle22pp\rangle$. This generalizes the discussion of \cite{Castro:2024cmf}, as in our case these operators also get $1/N$ contributions from the tree gluon exchange $M_F^2$. We then take the flat space limit \cite{Penedones:2010ue} of \eqref{MR2} and find a precise match with the 8d flat space tree graviton exchange \cite{Behan:2024vwg}
   \es{AR22}{
\cA^{ABCD}_R&=-32\bm\Delta\pi^7\delta^{AB}\delta^{CD}\int\frac{d^2p_\perp}{(2\pi^2)}\frac{1}{-s+p_\perp^2}+\text{crossed}\\
&=-8\bm\Delta\pi^6(\log (-s)+\dots)\delta^{AB}\delta^{CD}+\text{crossed}\,,
}
 where the logs come from the integral over the transverse momenta $p_\perp$, since the graviton probes the full 10d spacetime.

Finally, for $\langle2222\rangle$ in the $D_4$ theory, integrated constraints from derivatives of the mass deformed sphere free energy as computed from supersymmetric localization can be used to fix $M_F^4$ \cite{Chester:2022sqb,Behan:2023fqq}.\footnote{For the other theories one can similarly try to compute the mass deformed sphere free energy using Seiberg-Witten curves \cite{Behan:2024vwg}, but so far this has only been successful for the $\log N$ term.}$^,$\footnote{For other papers that use integrated constraints to fix correlators, see \cite{Binder:2019jwn,Binder:2021cif,Alday:2021ymb,Binder:2018yvd,Binder:2019mpb,Chester:2023qwo,Chester:2021aun,Chester:2020dja,Chester:2019pvm,Chester:2019jas,Alday:2021vfb,Alday:2022rly,Chester:2024bij,Chester:2024esn,Alday:2024yax,Chester:2023ehi,Alday:2023pet,Dempsey:2024vkf,Billo:2023ncz,Brown:2024yvt,Alday:2024srr,Cavaglia:2023mmu,Cavaglia:2022qpg,Caron-Huot:2024tzr,Chester:2025ssu}.} The previous study \cite{Behan:2023fqq} could only fix the $\tau$-dependent part of $M_F^4$, since $M_R$ contributes at the same order of $1/N^2$ and was not known. Using our new formula for $M_R$, we completely fix $M_F^4$. We then extract all unambiguous CFT data at $1/N^2$, and take the flat space limit to give a prediction for $\cA_{F^4}$.\footnote{The $\tau$-dependent parts of $\cA_{F^4}$ were computed in \cite{Kiritsis:2000zi} and matched in \cite{Behan:2023fqq}, the novelty here is the $\tau$-independent terms.}

The rest of this paper is organized as follows. In Section \ref{sec:sugra} we consider the effective bulk theory of 10d supergravity coupled to 8d SYM on the sevenbranes, and use this compute the bulk couplings $\beta_{pqk_r}$. In Section \ref{sec:scattering} we discuss gluon scattering $\langle 22pp\rangle$ on $AdS_5\times S^3$, including the graviton exchange term and mixing between the graviton modes and double traces. We conclude in Section \ref{sec:conclusion} with a review of our results and a discussion of future directions. Technical details of the calculations are given in the various Appendices.

\section{Extremal couplings in F-theory on AdS}\label{sec:sugra}
	We will be interested in holographic backgrounds sourced by $N$ D3-branes probing 7-brane singularities \cite{Morrison:1996na,Morrison:1996pp,Vafa:1996xn}. The 7-brane setups with a holographic SCFT dual have a constant axio-dilaton, for which there has been a classification according to the associated ADE-singularities \cite{Sen:1996vd,Dasgupta:1996ij}. In Table \ref{Tab: D7 brane singularities} we summarise the possible singularities and the associated values of the axio-dilaton $\tau = C_0 + \rmi \rme^{-\Phi}$, with an unconstrained value in the $D_4$ case. Upon backreacting the D3-branes all the listed 7-brane backgrounds have a metric that can be written as
	\begin{equation}
		\rmd s^2 =  L^2 \left( \rmd s_{\text{AdS}_5}^2 + \rmd \theta^2 + \cos^2 \theta\, \rmd \phi^2 + \sin^2 \theta \, \rmd s_{S^3}^2 \right)\,,
	\end{equation}
	where the coordinate ranges are $0 \leq \theta \leq \pi/2$ and $ 0 \leq \phi \leq 2\pi/\bm\Delta$. The associated volume of the internal manifold, which we will denote with $M_5$, is thus $\text{vol}_{M_5} = \text{vol}_{S^5}/\bm\Delta$. The value of $\bm\Delta$ is fixed in terms of the particular 7-brane singularity, and we have listed them in the second row of Table \ref{Tab: D7 brane singularities}. In the dual field theory this parameter corresponds to the conformal dimension of the lowest Coulomb branch operator \cite{Aharony:2007dj} as discussed in the introduction. The backgrounds additionally have non-trivial values for the five-form flux
	\begin{equation}
		F_5 = \rmd C_4 = 4 L^4 (1 + \star)\text{vol}_{M_5}\,,\qquad \text{with}\qquad \frac{1}{(2\pi \ell_s)^4 g_s} \int_{M_5} F_5 = N\,.
	\end{equation}
	The quantisation of $F_5$ relates the field theory and string theory parameters such that
\es{dictionary}{
		(L/\ell_s)^4 = 4\pi \bm\Delta g_s N\,.
}
	Although locally the metric on the internal space is simply that of $S^5$ the 7-branes break its isometries down to
	\begin{equation}
		\text{SO}(6)\rightarrow \text{SU}(2)_L \times \text{SU}(2)_R \times  \text{U}(1)_R\,,
	\end{equation}
	where $L$ and $R$ refer to the global flavor- and R-symmetries in the dual field theory, on top of the flavor symmetry $G_F$ arising from the 7-brane singularities. The backgrounds summarised above are solutions to the type IIB supergravity action
	\begin{equation}\label{Eq: IIB action}
		S_{\text{IIB}} = \frac{1}{2\kappa^2} \int \rmd^{10}X \sqrt{g} \, \left( R + \frac{1}{5!} F_5^2\right) \,,\quad \frac{1}{2\kappa_{}^2} = \frac{2\pi}{(2\pi \ell_s)^8 g_s^2} \equiv \frac{2\pi}{(2\pi \ell_p)^8}\,,
	\end{equation}
	and the 7-branes provide additional degrees of freedom which are effectively described by the standard DBI and WZ actions
	\begin{equation}
		S_{7} = T_{7} \left( \int \rmd^8 \hat X \sqrt{\text{det}(\hat g + 2\pi \ell_s^2 F)} +  \int \hat C_4 \wedge \rme^{2\pi \ell_s^2 F}\right) \,, \quad T_{7} =\frac{2\pi}{(2\pi \ell_s)^8 g_s}\,,
	\end{equation}
	where the hats denote pullbacks from the background onto the worldvolume of the branes and $F$ is the worldvolume flux.\footnote{We note that on the 7-brane singularities of interest the two-form gauge potentials are projected out.} These 7-branes sit at the fixed point locus $\theta = \pi/2$ and wrap AdS$_5$ as well as the $S^3 \subset M_5$.
	\subsection{Central charges}
	Before moving on to the cubic couplings we first compute the five-dimensional effective couplings that arise from reducing the ten-dimensional theory over the internal geometry and check that they are directly related to the central charges in the dual field theory \cite{Aharony:2007dj}. The effective five-dimensional gravitational coupling equals to
	\begin{equation}
		S_{\text{IIB}} = \frac{1}{2\kappa^2_5}  \int_{} \rmd^{5}x  \sqrt{g} R_{} + \ldots \,,\quad \text{where} \quad \frac{1}{2\kappa^2_5} = \frac{\text{vol}_{M_5}}{2\kappa^2} = \frac{\bm\Delta N^2}{8\pi^2  L^3}\,.
	\end{equation}
	The standard holographic dictionary then dictates that the two point function of the canonically normalised stress-tensor equals \cite{Liu:1998bu,Freedman:1998tz,Penedones:2016voo}
	%
	%
	%
	\begin{equation}\label{Eq: def of CT}
		\langle T_{\hat{\mu}\hat{\nu}}(z) T_{\hat{\rho}\hat{\sigma}}(0)\rangle =\, \frac{c}{640\pi^2 z^8} \left( I_{\hat{\mu}(\hat{\rho}}I_{\hat{\sigma})\hat{\nu}} - \text{trace}  \right) \,,\qquad   c \approx  \frac{\pi^2 L^3 }{\kappa^{2}_5} =  \frac{\mathbf{\Delta} }{4}\, N^2\,,
	\end{equation}
	%
	with $I_{\hat{\mu}\hat{\nu}} = \delta_{\hat{\mu}\hat{\nu}} - 2 \frac{z_{\hat{\mu}} z_{\hat{\nu}}}{z^2}$, and we have only computed the leading large $N$ contribution to the conformal anomaly $c$ \cite{Aharony:2007dj}.\footnote{The ten-dimensional coordinates are listed with $x_M$, while the AdS$_5$ and $M_5$ coordinates are denoted as $x_\mu$ and $y_a$ respectively and when the coordinates are pulled back onto the worldvolume of the 7-branes we put hats on them, e.g. $\hat y_{\hat a}$, the dual field theory coordinates are denoted with $z_{\hat{\mu}}$, and finally the indices $A,B,\ldots$ denote adjoint indices of the flavor group $G_F$.} 
	Similarly, the gauge coupling on the brane is found by expanding the DBI action to quadratic order and reducing this to five dimensions as well
	\begin{equation}
		S_{7} \approx  -\int \rmd^5  x \,\frac{1}{4 \text{g}_{5d}^2} (F_{\mu\nu}^A)^2 + \ldots\,,\quad \text{where}\quad \frac{1}{4 \text{g}_{5d}^2} =  \frac{N \bm\Delta }{8\pi^2 L}\,,
	\end{equation}
	where we have normalised the generators of the flavor group as
	\begin{equation}\label{Eq: normalisation of flavor group}
		\text{tr}\, T^A T^B = 2 \delta^{AB}\,.
	\end{equation}
	Again applying the standard holographic dictionary we find that the two point function of the flavor symmetry currents equals
	\begin{equation}
		\langle J^A_{\hat{\mu}}(x) J^B_{\hat{\nu}}(0)\rangle =\, \frac{ 3{\bf k}}{2\pi^2} \frac{1}{z^6}\delta^{AB} I_{\hat{\mu}\hat{\nu}}\,,\qquad  \mathcal {\bf k} = \frac{8 \pi^2 L }{\text{g}_{5d}^2} =  2\mathbf{\Delta} N\,,
	\label{eq: flavor central charge}
	\end{equation}
	where the flavor central charge $\bf k$ is in fact large $N$ exact \cite{Aharony:2007dj}.

%
	\subsection{Cubic couplings}
	Now we move on to computing the cubic couplings between two scalar KK-modes coming from type IIB supergravity, and gluon modes coming from the 7-branes. Although the internal geometry depends on the choice of 7-brane singularity, for the KK-analysis these differences will play only a minor role and we can use most of the results in the context of AdS$_5 \times S^5$ \cite{Kim:1985ez,Lee:1998bxa,Arutyunov:1998hf}. In particular, when expanding the IIB action in KK-modes one finds that there are scalar degrees of freedom coming from both the metric and the four-from gauge potential which have to be diagonalised. This diagonalisation procedure can be done through a careful study of the type IIB equations of motion \cite{Kim:1985ez}, which we do not repeat here. Instead, we summarise the vital results in the aforementioned references and apply them to our cases of interest. 
	
	After gauge fixing the metric fluctuations one finds that the scalar modes come in the following form \cite{Kim:1985ez}
	\begin{equation}\label{Eq: metric fluctuations}
	\begin{aligned}
		\delta g_{ab} = h_{ab} = h'_{ab} + \frac{h}{5} g_{ab} \,,\quad \text{with}\quad g^{ab}h'_{ab} = 0\,,\quad h'_{ab} = h'_{ab}\,,\\
		\delta g_{\mu\nu} = h_{\mu\nu} = h'_{\mu\nu} - \frac{3h}{25} g_{\mu\nu}\,,\quad \text{with}\quad g^{\mu\nu}h'_{\mu\nu} = 0\,,\quad h'_{\mu\nu} = h'_{\nu\mu}\,,
	\end{aligned}
	\end{equation}
	and similarly the scalar fluctuation in the four-form equals
	\begin{equation}
		\delta C_4 = ((\star \rmd)_{X_{5}} - (\star \rmd)_{\text{AdS}_5} )a\,.
	\end{equation}
	These fluctuations can subsequently be expanded into harmonics on the internal manifold, of which we will only list the scalar ones:
	\begin{equation}\label{Eq: KK fluctuations on S5}
	\begin{aligned}
		h(x,y) =& \sum_k h_{k}(x) \mathcal Y^{I_k}(y) \,,\\
		h'_{(\mu\nu)}(x,y) =& \sum_k h'_{(\mu\nu),k}(x) \mathcal Y^{I_k}(y)\,,\\
		a(x,y) =& \sum_k a_k(x) \mathcal Y^{I_k}(y)\,,
	\end{aligned}
	\end{equation}
The scalar harmonics $\mathcal Y^{k_p}$ can be readily constructed as symmetric traceless polynomials of some ambient $\R^6$ coordinates $y_{1,\ldots , 6}$. As mentioned already we are particularly interested in harmonics of rank $k_p$ that transform in the spin-representation $(\tfrac{p}{2}-1,\tfrac{p}{2}-1)_0$ of $\text{SU}(2)_L \times \text{SU}(2)_R \times  \text{U}(1)_R \subset \text{SO}(6)$. To explicitly keep track of this additional label $p$ we will therefore from now on denote the rank of the harmonics as $k_p$. With respect to the $S^5$ and $S^3$ Laplacians these harmonics have eigenvalues
	\begin{equation}
		\Delta_{S^5} \mathcal Y^{I_{k_p}} = -k_p (k_p + 4)\,,\quad \text{and} \quad \Delta_{S^3} \mathcal Y^{I_{k_p}} = -p(p - 2)\,,
	\end{equation}
	where both $k_p$ and $p$ are non-negative integers, and\footnote{Note that we also have modes $k_2=0$ and $k_3=1$, but these are pure gauge in the bulk and so not physical.}
	\begin{align}
  k_p = \left\{ \begin{aligned}
  	\,\, p, p+2, p+4,\dots		\qquad & p=2,3			\\
  	\,\, p-2,p,p+2,\dots 		\qquad & p>3 \ .
  \end{aligned} \right.
\label{eq: kp range main text}
\end{align}
	In terms of the symmetric traceless polynomials these harmonics are explicitly constructed as
	\begin{equation}\label{Eq: harmonics Y main text}
		\mathcal Y^{I_{k_p}} = T_{i_1 i_2 \ldots i_{p-2}}\, y_{i_1} y_{i_2} \cdots y_{i_{p-2}} \, {}_2 F_1 \left(\frac{p -2 - k_p}{2}, \frac{p + 2 + k_p}{2}, 1,\rho^2\right)\, ,
	\end{equation}
	where $T_{i_1 i_2 \ldots i_{p-2}}$ is a symmetric traceless tensor in $\text{SO}(4) \cong \text{SU}(2)_L \times \text{SU}(2)_R$, the embedding coordinates $y_{i=1,\ldots,4}$ are restricted to the associated three-sphere, and $\rho^2 = y_5^2 + y_6^2$. We refer the reader to Appendix \ref{app: cubic couplings} for the details on these harmonics and only mention here that the hypergeometric function reduces to a finite polynomial given the fact that its first argument is a negative integer.
		
	The type IIB equations of motion dictate a mixing between the fluctuations $a_{k_p}(x)$ and $h_{k_p}(x)$, which can be diagonalised by taking the following linear combinations
	\begin{equation}\label{Eq: metric and four form in terms of s and t}
	\begin{aligned}
		h_{k_p}(x) = 10 k_p s_{k_p}(x) + 20(k_p + 4)t_{k_p}(x)\,, \quad a_{k_p}(x) = t_{k_p}(x) - s_{k_p}(x) \,,
	\end{aligned}
	\end{equation}
	which consequently have the following equations of motion
	\begin{equation}
		(\square- k_p(k_p-4)) s_{k_p} =  (\square - (k_p + 8)(k_p + 4)) t_{k_p} = 0\,.
	\end{equation}
	The modes $s_{k_p}$ are dual to scalar superconformal primary operators on the boundary, while the modes $t_{k_p}$ are dual to superdescendants. We only have interest in the superprimary modes and will set $t_{k_p} = 0$ by hand in what follows. The quadratic action of the superprimary modes then reduces to
	\begin{equation}
	\begin{aligned}
		S^{(2)}(s_{k_p}) =& \frac{1}{2\kappa^2}  \frac{64k_p(k_p-1)}{(k_p + 1)(p-1)}\frac{\pi^3}{\bm\Delta} \int_{} d^5x \sqrt{g} \, \left( (\partial_\mu s_{k_p})^2 + m_{{k_p}}^2 s_{k_p}^2 \right)\,,
	\end{aligned}
	\end{equation}
	where $m_{{k_p}}^2 = k_p(k_p-4)$ and consequently $\Delta_{s_{k_p}} = k_p$. The prefactor in the equation above arises from integrating the harmonics in \eqref{Eq: harmonics Y main text} over the five sphere. We emphasise that these scalar modes are in non-trivial representations of $\text{SU}(2)_L \times \text{SU}(2)_R \subset \text{SO}(6)$, we have however omitted the associated indices for notational simplicity. The full form, including an explicit contraction of all indices can be found in Appendix \ref{app: cubic couplings}. 
	
	An important fact that needs to be kept in mind is that the IIB equations of motion relate the fluctuations between the spin-2 modes and the scalar modes $s_{k_p}(x)$:
	\begin{equation}\label{Eq: spin 2 constraint}
		h'_{(\mu \nu),k_p} =H_{(\mu\nu),k_p}' + \frac{4 \nabla_{(\mu} \nabla_{\nu)} s_{k_p}}{k_p + 1} \,,
	\end{equation}
	which does not alter the quadratic action of the scalars, but will be of importance when computing the cubic couplings. We will only be interested in couplings of scalar fields, and consequently will turn off the tensor fluctuations: $H'_{(\mu\nu)}=0$. The scalar couplings of interest arise from expanding the action of the 7-branes to cubic order. Only keeping the relevant terms contributing to the masses of the gluon modes and the cubic couplings of interest we find that
	\begin{equation}\label{Eq: cubic expansion brane action}
		\begin{aligned}
			S_{\text{brane}}^{(3)} = -2\pi^2 \ell_s^4  T_{7}\int \Big[ \rmd^8 \hat X \sqrt{\hat g}\Big(& F^{A,\hat M\hat N} F_{\hat M\hat N}^A - 2 F^{A,\hat M\hat N} F_{\hat N\hat P}^A \hat h^{\hat P}_{\phantom{P}\hat M} 
			\Big) - 2\hat C_4 \wedge F^A \wedge F^A  \Big]\,,
		\end{aligned}
	\end{equation}
	where the worldvolume flux $F^A = (\rmd A^A - \rmi  f_{ABC} A^B \wedge A^C )T^A$ is non-abelian with structure constants $f_{ABC}$. The gauge field is consequently expanded in harmonics on the three sphere which the 7-branes wrap\footnote{Since there can not be any confusion about the pulled back directions along AdS$_5$ we will not denote these coordinates with hats on them, or their corresponding indices.} 
	\begin{equation}
		A^A_\mu = \sum\limits_{I} a^{A,I}_\mu( x) \hat{\mathcal Y}^{I}(\hat y) \,,\quad A^A_{\hat{a}} = \varphi^A( x) \cdot \mathcal X_{\hat{a}}( \hat y)\,,
	\end{equation}
	where $\hat{\mathcal Y}$ are scalar, and $\mathcal X_{\hat{a}}$ are vector harmonics. We are only interested in scalar modes in AdS dual to chiral superprimary operators in the dual CFT, and will consequently put $a^{A,I}_\mu = 0$. The inner product in $A_{{\hat{a}}}^A$ is determined by the particular representation of $\mathcal X_{\hat{a}}(\hat y)$, with respect to the $\text{SU}(2)_L \times \text{SU}(2)_R \times \text{U}(1)_R$ isometries, i.e. the spin-labels $(j_L,j_R)_r$. There are two particular towers of vector modes on $S^3$ which transform as \cite{Aharony:1998xz,Kruczenski:2003be}
	\begin{equation}
		\mathcal X_{\hat{a}}^{p}(\hat y) :\,  \left(\tfrac{p}{2}-1,\tfrac{p}{2} \right)_0 \qquad \text{and} \qquad  \tilde{\mathcal X}_{\hat{a}}^{p}(\hat y) :\, \left(\tfrac{p}{2},\tfrac{p}{2}-1 \right)_0 \,,
	\end{equation}
	where $p = 2,3,\ldots$ . The dual scalar operators have respectively the scaling dimensions $\Delta_{p} = p$ and $p+4$, and combine into $\mathcal N=2$ multiplets. We will only keep track of the superprimary operators, whose vector modes are given by $\mathcal X_{\hat{a}}^{p}(\hat y)$. These vector harmonics are constrained by the following differential equations
	\begin{equation}\label{Eq: constraints vector harmonics}
	\begin{aligned}
	 {\nabla}^{\hat{b}} {\nabla}_{\hat{b}} \mathcal X_{\hat{a}}^{p} = (2 - p^2) \mathcal X_{\hat{a}}^{p}, \quad 
    {\epsilon_{{\hat{a}}}}^{\hat b\hat c} {\nabla}_{\hat{b}} \mathcal X_{\hat{c}}^{p} = -p \mathcal X_{\hat{a}}^{p}, \quad
   	{\nabla}^{\hat{a}} \mathcal X_{\hat{a}}^{p} = 0\,.
	\end{aligned}		
	\end{equation}
	Using these constraints the quadratic action for the gluon modes $\varphi^{A}_{(p)}$ can be easily computed from the Yang-Mills term in \eqref{Eq: cubic expansion brane action} and equals
	\begin{equation}
	\begin{aligned}
		S^{(2)}(\varphi_{p}) =&\, - \frac{4\pi^4 \ell_s^4 T_7}{p-1}  \int \rmd^5  x \sqrt{ g} \Big(\, (\partial_{{\mu}} \varphi^{A}_{(p)})^2  + m_{p}^2 (\varphi^{A}_{(p)})^2 \Big)\,,\quad m_{p}^2 = p(p-4)\,,
	\end{aligned}
	\end{equation}
	where we have integrated the vector harmonics over the internal three sphere. Just as we have done for the metric fluctuations above, we have suppressed all indices on $\varphi^A_{(p)}$ that are related to the non-trivial representation under $\text{SU}(2)_L \times \text{SU}(2)_R$, and relegated these details to Appendix \ref{app: cubic couplings}.
	
	Finally, we are interested in the cubic couplings between two gluon modes ($\varphi_{(p)}^A$ and $\varphi_{(q)}^A$) and the tower of graviton modes ($s_{(k_r)}$). This coupling is computed from the second and last term in \eqref{Eq: cubic expansion brane action}. After canonically normalising all the fields, making use of \eqref{Eq: metric and four form in terms of s and t}, \eqref{Eq: spin 2 constraint}, and \eqref{Eq: constraints vector harmonics}, and integrating over the internal manifold we find that the cubic action, including the gluon-gluon-graviton coupling, equals
	%
	%
	%
	\begin{equation}
	\begin{aligned}
		S^{(3)} = \, \int \rmd ^5 x \sqrt{g}&  \bigg(\,\sum_{p=2 }^\infty \left(-\frac{1}{2}(\partial_{{\mu}} \varphi^{A}_{(p)})^2 -\frac{1}{2} m_{p}^2 (\varphi^{A}_{(p)})^2\right)+\sum_{p=2}^\infty\sum_{k_p}\left( -\frac{1}{2} (\partial_\mu s_{(k_p)})^2  -\frac{1}{2} m_{k_p}^2 s_{(k_p)}^2\right) \nn\\
		&\qquad +\sum_{p,q=2}^\infty \sum_r \sum_{k_r} \frac{1}{2} \beta_{pq k_r}  s_{(k_r)}\varphi_{(p)}^A \varphi_{(q)}^A \bigg)\,,
	\end{aligned}
	\end{equation}
	where $r$ runs over the values allowed by $SU(2)_L\times SU(2)_R\times U(1)_R$ invariance, namely
	\begin{align}
  r=|p-q|+2, |p-q|+4,\dots, p+q-2
\end{align}
and for fixed $p$, $k_p$ runs over the range in \eqref{eq: kp range}.
	The cubic coupling equals
	\begin{equation}
	\begin{aligned}
		\beta_{pqk_r} = &\frac{\pi}{2 N} 
\sqrt{\frac{(p - 1)(q - 1)(r - 1)}{2 \bm\Delta k_r (k_r - 1)(k_r + 1)}}
(k_r^2 - (p - q)^2)(k_r + p + q - 4)(k_r + p + q - 2) \\
&\times \frac{
\Gamma(p - 1)\Gamma(q - 1)\Gamma(r - 1)
}{
\Gamma\left(\frac{p + q - r}{2}\right)
\Gamma\left(\frac{p - q + r}{2}\right)
\Gamma\left(\frac{q - p + r}{2}\right)
\Gamma\left(\frac{p + q + r - 2}{2}\right)
} \,.
	\end{aligned}
	\label{eq: coupling final answer}
	\end{equation}
Note that this coupling is between three half-BPS multiplets when $k_r = r-2$. Assuming this, it is furthermore extremal or super-extremal precisely if $r=|p-q|+2$, when it is extremal. It is then straightforward to see that that in this case, $\beta_{pqk_r}=0$. This in particular substantiates the earlier claim, that for the coupling $\beta_{2pk_p}$ of relevance to our 4-point calculation, we have $\beta_{2pk_p}\neq 0$ only for $k_p=p,p+2,\dots$.

	From the cubic coupling one can directly predict the field theory OPE coefficient $\lambda_{pqk_r}$ between two gluon KK-modes and a graviton KK-mode using \eqref{3point} to find that the leading $1/N$ contribution equals
	%
	\begin{equation}\label{beta}
	\lambda_{pq k_r}=\frac{1}{N} \sqrt{\frac{r - 1}{\bm\Delta k_r (k_r + 1)}} 
\frac{
\Gamma\left(\frac{2 + k_r + p - q}{2}\right)
\Gamma\left(\frac{2 + k_r - p + q}{2}\right)
\Gamma\left(\frac{p + q - k_r}{2}\right)
\Gamma\left(\frac{k_r + p + q}{2}\right)
\Gamma(r - 1)
}{
\Gamma(k_r)
\Gamma\left(\frac{p + q - r}{2}\right)
\Gamma\left(\frac{p - q + r}{2}\right)
\Gamma\left(\frac{q - p + r}{2}\right)
\Gamma\left(\frac{p + q + r - 2}{2}\right)
}+O(1/N^2).
	\end{equation}
	We note that the relation between the cubic couplings and the OPE coefficients is only valid for non-extremal couplings. It is clear from the result above, however, that such (super-)extremal couplings do occur, making the coefficient diverge. The resolution is that in those (super-)extremal cases one first has to diagonalise a mixing problem between single and double trace operators, after which the non-extremal OPE coefficients can be reliably computed. We will explicitly perform such a unmixing procedure in the following section.
	
	Finally, as a consistency check of our result we explicitly determine the OPE coefficient in the case that $p=q$, $r=2$ and $k_2=2$, in which case it matches the expected OPE coefficient $\lambda_{pp(k_2=2)}$ between two scalars and the stress tensor supermultiplet \cite{Chang:2017xmr} for our theories \cite{Alday:2021odx,Aharony:2007dj}:
	\es{check}{
	 \lambda_{pp(k_2=2)}=\frac{p}{N\sqrt{6\mathbf{\Delta}}}+O(1/N^2)\,.
	}

\section{Gluon scattering in $AdS_5\times S^3$}\label{sec:scattering}
We will now use the bulk couplings determined in the previous section to compute the contribution of graviton exchange to gluon scattering $\langle22pp\rangle$ on sevenbranes in $AdS_5\times S^3$. This scattering is dual to certain $1/N^2$ corrections to the moment map four-point function in certain 4d $\mathcal{N}=2$ CFTs with flavor group $G_F=A_1, A_2, D_4, E_6,E_7,E_8$. We begin by reviewing the large $N$ expansion of this correlator to order $1/N^2$, as fixed from the analytic bootstrap in \cite{Alday:2021odx,Alday:2021ajh,Behan:2023fqq,Behan:2024vwg,Huang:2023ppy}. We then discuss how to compute the graviton exchange term, and how we can extract CFT data after taking into account the mixing due to extremal couplings \cite{Castro:2024cmf}. Finally, we specialize to the $G_F=D_4$ theory and $\langle2222\rangle$, where we can use localization constraints to completely fix the correlator to order $1/N^2$ and extract all unambiguous CFT data.

\subsection{Setup}
\label{setup}

The 4d $\cN=2$ CFTs we consider have R-symmetry $SU(2)_R\times U(1)_R$ and flavor symmetries $SU(2)_L\times G_F$. We consider half-BPS multiplets dual to the $p$-th gluon KK mode, whose superprimary is the Lorentz scalar  $\phi_p^A(y,\bar y,x)$ with scaling dimension $\Delta=p$, which transforms in the isospin $\frac p2-1$ irrep of $SU(2)_L$ with spinor polarizations $\bar y$, the isospin $\frac p2$ irrep of $SU(2)_R$ with spinor polarizations $y$, and the adjoint $\mathfrak{g}$ of $G_F$ with index $A=1,\dots,\dim(G_F)$. Then, $\phi^A_p$ is the CFT operator dual to the bulk field $\varphi^A_p$. The conformal and global symmetries restrict $\langle \phi_2 \phi_2 \phi_p \phi_p \rangle$ (denoted as $\langle22pp\rangle$) to be
\es{phiExp1}{
&\langle \phi_2^A(y_1,x_1) \phi_2^B(y_2,x_2) \phi_p^C(y_3,\bar y_3,x_3) \phi_p^D(y_4,\bar y_4,x_4) \rangle \\
&\hspace{60mm}= \frac{\langle y_1,y_2\rangle^2 \langle y_3,y_4\rangle^p \langle \bar y_3,\bar y_4\rangle^{p-2}}{x_{12}^4x_{34}^{2p}}\sum_{\bf r\in\mathfrak{g} \otimes \mathfrak{g}}G_{\bf r}(U,V;w)P_{\bf r}^{ABCD}\,,\\
&\langle \phi_2^A(y_1,x_1) \phi_p^B(y_2,x_2) \phi_2^C(y_3,\bar y_3,x_3) \phi_p^D(y_4,\bar y_4,x_4) \rangle\\
&\hspace{60mm} = \frac{\langle y_1,y_2\rangle^2 \langle y_3,y_4\rangle^2\langle y_2,y_4\rangle^{p-2} }{x_{12}^4x_{34}^{4}x_{24}^{2(p-2)}}\sum_{\bf r\in\mathfrak{g} \otimes \mathfrak{g}}\widetilde G_{\bf r}(U,V;w)P_{\bf r}^{ABCD}\,,
}
where for later convenience we define both $G_{\bf r}$ in the $\langle22pp\rangle$ configuration and $\widetilde G_{\bf r}$ in the $\langle2p2p\rangle$ configuration, while the explicit projectors are written in Appendix A of \cite{Behan:2024vwg} in terms of the dual coxeter number $h^\vee$. Superconformal symmetry can be solved by writing $G_{\bf r}(U,V;w)$ in terms of a reduced correlator $\mathcal{G}_{\bf r}(U,V)$ \cite{Dolan:2001tt}
\es{ward4dN2}{
G_{\bf r}(U,V;w)=\frac{z(w-\bar z)f_{r}(\bar z)-\bar z(w-z)f_{r}(z)}{w(z-\bar z)}+\left ( 1-\frac{z}{w} \right ) \left ( 1-\frac{\bar z}{w} \right )\mathcal{G}_{\bf r}(U,V)\,,
}
where $U=z\bar z\,, V=(1-z)(1-\bar z)\,, w=\frac{\langle y_1,y_2\rangle\langle y_3,y_4\rangle}{\langle y_1,y_3\rangle\langle y_2,y_4\rangle}$ and similarly for $\widetilde G_{\bf r}$ in terms of $\tilde f_{\bf r}$ and $\widetilde{\mathcal G}_{\bf r}$. By taking the OPE twice, we can expand $f$ and $\mathcal{G}$ in terms of conformal blocks as \cite{Beem:2013sza,Chester:2025ssu}\footnote{We denote the short OPE coefficients squared by the dimension of their superprimary, which is shifted by two relative to the notation in \cite{Chester:2022sqb}.}
\es{blockExpSQCD}{
\mathcal{G}_{\bf r}(U,V)&=U^{-1}\Big[\sum_{\ell}\sum_{\Delta\geq\ell+2}\lambda^{(2,p)}_{\Delta,\ell,\bf r}g^{0,0}_{\Delta+2,\ell}(U,V)+\sum_{\ell}\lambda_{\ell+3,\ell+1,\bf r}^{(2,p)}g^{0,0}_{\ell+5,\ell+1}(U,V)+\lambda_{2,0,\bf r}^{(2,p)}g^{0,0}_{4,0}(U,V)\Big]\,,\\
f_{\bf r}(z)&=\sum_{\ell}\lambda_{\ell+3,\ell+1,\bf r}^{(2,p)}k^{0,0}_{2\ell+6}(z)+\lambda_{2,0,\bf r}^{(2,p)} k^{0,0}_{4}(z)+\delta_{\bf r,{\bf 1}}\dim(G)-\delta_{\bf r,\text{Adj}}\frac{4h^{\vee}}{{\mathbf k}}k^{0,0}_2(z)-\delta_{\bf r,{\bf1}}\frac{p\,\text{dim}(G)}{12c}k^{0,0}_4(z)\,,\\
}
where $\lambda^{(2,p)}_\cO\equiv \lambda_{22\cO} \lambda_{pp\cO}$, we have even/odd spin $\ell$ blocks for the irreps $r$ in the symmetric/antisymmetric product of the adjoint, and the 4d blocks and lightcone blocks are defined as
\es{4dblock}{
 g^{\Delta_{12},\Delta_{34}}_{\Delta,\ell}(U,V) &=\frac{z\bar z}{z-\bar z}(k^{\Delta_{12},\Delta_{34}}_{\Delta+\ell}(z)k^{\Delta_{12},\Delta_{34}}_{\Delta-\ell-2}(\bar z)-k^{\Delta_{12},\Delta_{34}}_{\Delta+\ell}(\bar z)k^{\Delta_{12},\Delta_{34}}_{\Delta-\ell-2}( z))\,,\\
k^{\Delta_{12},\Delta_{34}}_h(z)&\equiv z^{\frac h2}{}_2F_1(h/2+\Delta_{12}/2,h/2-\Delta_{34}/2,h,z)\,.
}
The flavor central charge and conformal anomaly for our theories are
\es{centralcharges}{
c=\frac{\bm\Delta}{4}\,N^2+\frac{3(\bm\Delta-1)}{4}\,N-\frac{1}{12}\,,\quad
{\mathbf k}=2\,\bm\Delta\,N\,.
}
The protected OPE coefficients can be trivially computed from the free theory, and the only one we will make use of is \cite{Beem:2013sza}
\es{OPEc}{
\lambda^{(2,p)}_{2,0,\bf{1}}=2\delta_{2,p}+\frac{p\,\text{dim}(G_F)}{12c}-\frac{4h^\vee}{{\mathbf k}}\,.
}
We can similarly expand $\widetilde{\mathcal{G}}_{\bf r}$ in terms of blocks to get \cite{Huang:2023ppy}
\es{blockExpSQCD2}{
\widetilde{\mathcal{G}}_{\bf r}(U,V)&=U^{-\frac{p}{2}}\sum_{\ell}\sum_{\Delta\geq\ell+p}\lambda_{2p(\Delta,\ell,\bf r)}^2g_{\Delta+2,\ell}^{2-p,2-p}(U,V)+\dots\,,
}
where the ellipses denotes short blocks that we will not use, and we now have both even and odd spins for each irrep.

For the $1/N$ expansion of the reduced correlator, we find it convenient to work in Mellin space as defined by
\es{mellin}{
\mathcal{G}_\text{conn}^{ABCD}(U,V)&=\int \frac{ds dt}{(4\pi i)^2}U^{\frac s2}V^{\frac{t-p-2}{2}} M^{ABCD}(s,t) \Gamma\Big[2-\frac s2\Big]\Gamma\Big[p-\frac s2\Big]\Gamma\Big[\frac{2+p-t}{2}\Big]^2\Gamma\Big[\frac{2+p-u}{2}\Big]^2\,,\\
\widetilde{\mathcal{G}}_\text{conn}^{ABCD}(U,V)&=\int \frac{ds dt}{(4\pi i)^2}U^{\frac {s+2-p}{2}}V^{\frac{t-p-2}{2}} M^{ABCD}(s,t) \Gamma\Big[\frac {2+p-s}{2}\Big]^2\Gamma\Big[\frac{2+p-t}{2}\Big]^2\Gamma\Big[2-\frac{u}{2}\Big]\Gamma\Big[p-\frac{u}{2}\Big]\,,\\
}
where $s+t+{u}=2(1+p)$, the two integration contours include all poles in $s,t$ but not $u$, and we define the connected correlator by subtracting the disconnected correlator as
\es{connected}{
\mathcal{G}_\text{conn}^{ABCD}(U,V)&=\mathcal{G}^{ABCD}(U,V)-(\delta^{AB}\delta^{CD}+\delta_{2,p}U\delta^{AC}\delta^{BD}+\delta_{2,p}\frac{U}{V^2}\delta^{AD}\delta^{BC})\,,\\
\widetilde{\mathcal{G}}_\text{conn}^{ABCD}(U,V)&=\widetilde{\mathcal{G}}^{ABCD}(U,V)-(\delta_{2,p}\delta^{AB}\delta^{CD}+U^{\frac{p}{2}}\delta^{AC}\delta^{BD}+\delta_{2,p}\frac{U}{V^2}\delta^{AD}\delta^{BC})\,.\\
}
The $1/N$ expansion of $\langle22pp\rangle$ is then fixed by crossing symmetry and analyticity (i.e. the analytic bootstrap) to be \eqref{mellin_generalP}.
The tree-level gluon exchange term $M_{F^2}$ is \cite{Alday:2021odx}
\es{gluontree}{
M_{F^2}(s,t)=-\frac{4}{(p-2)!\bm\Delta}\left[\frac{\mathtt{c}_s}{(s-2)(u-p)}-\frac{\mathtt{c}_t}{(t-p)(u-p)}\right]\,,
}
where we define the tensors
\es{c_tree}{
\mathtt{c}_s^{ABCD}=f^{ABJ}f^{JCD}\,,\quad 
\mathtt{c}_t^{ABCD}=f^{ADJ}f^{JBC}\,,\quad 
\mathtt{c}_u^{ABCD}=f^{ACJ}f^{JDB}\,.
}
The one-loop gluon amplitude $M_{F^2|F^2}$ is \cite{Alday:2021ajh,Huang:2023ppy} 
\es{gluonloop}{
M_{F^2|F^2}(s,t)=-\frac{p}{4(p-2)!\bm\Delta^2}\left[\mathtt{d}_{st}\mathcal{B}^p_{st}+\mathtt{d}_{su}\mathcal{B}_{su}^p+\mathtt{d}_{tu}\mathcal{B}_{tu}^p\right]\,,
}
where the tensor structures are
\es{d_loop}{
\mathtt{d}_{st}^{ABCD}&=f^{IAJ}f^{JBK}f^{KCL}f^{LDI}\,,\\
\mathtt{d}_{su}^{ABCD}&=f^{IAJ}f^{JBK}f^{KDL}f^{LCI}\,,\\
\mathtt{d}_{tu}^{ABCD}&=f^{IAJ}f^{JCK}f^{KBL}f^{LDI}\,,
}
and the other coefficients can be written as the infinite sums
\es{Bs}{
\mathcal{B}^p_{st}&=\sum_{m,n=0}^\infty\frac{ (m-n) (m+2 n+3)-m (n+1) (p+1) (m+n+2)}{54 (m+n) (m+n+1) (m+n+2)(s-4-2m)(t-2-p-2n)}\,,\\
   \mathcal{B}^p_{su}&=\sum_{m,n=0}^\infty\frac{ (m-n) (m+2 n+3)-m (n+1) (p+1) (m+n+2)}{54 (m+n) (m+n+1) (m+n+2) (s-4-2m)(u-2-p-2n)}\,,\\
   \mathcal{B}^p_{tu}&=\sum_{m,n=0}^\infty\frac{ (m-n)^2-(m+1) (n+1) (p+1) (m+n)}{54 (m+n) (m+n+1) (m+n+2) (t-2-p-2m)(u-2-p-2n)}\,.\\
}
Note that these infinite sums are divergent, but the divergence is just a constant that can be absorbed into the definition of the contact term $M_{F^4}$. In a later section we will fix this constant for the special case of $p=2$, in which case we define the resummed and regularized $M_{F^2|F^2}$ in terms of 
\es{resummed}{
\mathcal{B}^2_{st} &= R_0(s, t) \left [ \psi^{(1)}(2 - \tfrac{s}{2}) + \psi^{(1)}(2 - \tfrac{t}{2}) - \left ( \psi(2 - \tfrac{s}{2}) - \psi(2 - \tfrac{t}{2}) \right )^2  \right ] \\
&+ R_1(s, t)\psi^{(0)}(2 - \tfrac{s}{2}) + R_1(t, s)\psi^{(0)}(2 - \tfrac{t}{2}) + \pi^2 R_2(s, t) + \frac{4}{27(s + t - 8)} -\frac{\gamma}{9}\,,
}
where the other coefficients are related by crossing as $\mathcal{B}^2_{st}=\mathcal{B}^2_{su}=\mathcal{B}^2_{tu}$. Here $\gamma$ is the Euler constant and we have the coefficient functions 
\es{Rs}{
R_0(s, t) &= \frac{3s^2t - 8s^2 + 3st^2 - 32st + 60s - 8t^2 + 60t - 96}{54(s + t - 8)(s + t - 6)(s + t - 4)} \,,  \\
R_1(s, t) &=- \frac{3s^2 + 3st - 26s - 10t + 48}{27(s + t - 8)(s + t - 4)}, \quad R_2(s, t) = -R_0(s, t)\,.
}
The contact terms $M^1_{F^4}$ and $M_{\log}$ are identical and take the form
\es{contact}{
M^1_{F^4,s}(s,t)=M_{\log,s}(s,t)=\mathtt{t}_1\,,\qquad M^1_{F^4,tu}(s,t)=M_{\log,tu}(s,t)=\mathtt{t}_2+\mathtt{t}_3\,,
}
where we define the tensors
\es{deltas}{
\mathtt{t}_1^{ABCD}&=\delta^{AB}\delta^{CD}\,,\quad
\mathtt{t}_2^{ABCD}=\delta^{AC}\delta^{DB}\,,\quad
\mathtt{t}_3^{ABCD}=\delta^{AD}\delta^{BC}\,,\\
}
while $M^i_{F^4}$ for $i\neq1$ only exist for the $D_4$ theory that we will discuss later. For $p=2$, crossing symmetry relates the $s$ and $t,u$ channels so we can simply define:
\es{p2log}{
p=2:\qquad M^1_{F^4}(s,t)=M_{\log}(s,t)=\mathtt{t}_1+\mathtt{t}_2+\mathtt{t}_3\,.
}
The coefficient $b_{\log}$ that multiplies $M_{\log}(s,t)$ for $p=2$ was computed in \cite{Behan:2023fqq,Behan:2024vwg} using supersymmetric localization and takes the form
\es{blog_fixed}{
b_{\log}=\frac{6}{\bm\Delta}\,.
}
The coefficients $b^i_{F^4}$ is unknown except for the case of $p=2$ for the $D_4$ theory, which will be discussed in subsection \ref{so8}. Finally, $M_R$ is the graviton exchange amplitude that we will compute next.

\subsection{Graviton exchange}
\label{grav}

The graviton exchange term $M_R$ for $\langle22pp\rangle$ in the direct channel includes contributions from the 4d $\cN=2$ stress tensor as well as an infinite tower of long multiplets in the singlet irrep of the flavor symmetry with scalar superprimary of dimension $k_2=2,4,6,\dots$. In the crossed channels, it gets contributions from the infinite tower of long multiplets $k_p=p+2,p+4,\dots$. The contribution of a scalar exchange diagram with scaling dimension $k_2$ and OPE coefficients $\lambda_{22k_2}\lambda_{ppk_2}$ to the reduced correlator takes the form \cite{Alday:2024yax,Alday:2021odx}
 \es{K}{
-\lambda_{22k_2}\lambda_{ppk_2}\sum_{m=0}\frac{1}{s-2m-k_2}\frac{4^{k_2 +1} \Gamma \left(\frac{k_2 +1}{2}\right) \Gamma
   \left(\frac{k_2 +3}{2}\right)}{\pi  (k_2/2)!^2 m! (1-m-k_2/2)!(p-1-m-k_2/2)!(m+k_2)!}\,,
   }
where the sum over $m$ corresponds to conformal descendants, and this expression has the same poles in $s$ as the corresponding block in \eqref{blockExpSQCD}.\footnote{The shift of scaling dimension by two and the overall power of $U^{-1}$ are taken into account in this formula.} Note that this sum is naively zero except for the case $k_2=2$, where it truncates to the $m=0$ term. Similarly, a scalar exchange diagram in the $\langle2p2p\rangle$ configuration corresponds to the block expansion in \eqref{blockExpSQCD2} and takes the form
 \es{K2}{
-\lambda_{2pk_p}^2\sum_{m=0}\frac{1}{s-2m-k_p}\frac{2  k_p!  (k_p+1)!}{ m!
    (k_p+m)! \Gamma
   (\frac{k_p}{2}-\frac{p}{2}+2)^2 \Gamma
   (\frac{k_p}{2}+\frac{p}{2})^2 \Gamma
   (1-\frac{k_p}{2}-m+\frac{p}{2})^2}\,,
   }
   which again is naively zero except for $k_p=p$ where it truncates to $m=0$.

However, writing $\lambda_{22k_2}\lambda_{ppk_2}$ and $\lambda_{2pk_p}^2$ in terms of the finite bulk coupling $\beta_{22k_2}\beta_{ppk_2}$ and $\beta_{2pk_p}^2$ using \eqref{3point} gives divergent factors for $k_2>2$ and $k_p>p$, respectively:
   \es{opeToBulk}{
 \lambda_{22k_2}\lambda_{ppk_2}&= \beta_{22k_2}\beta_{ppk_2}  \frac{\Gamma (2-k_2/2) \Gamma (k_2/2)^5 \Gamma (p-k_2/2) \Gamma (k_2/2+p-2)}{32 \pi ^2 \Gamma ( k_2)
   \Gamma ( k_2-1) \Gamma (p-1) \Gamma (p)} \,\\
    \lambda_{2pk_p}^2&= \beta_{2pk_p}^2 \frac{\Gamma \left(\frac{1}{2} (k_p-p+2)\right)^2 \Gamma
   \left(\frac{1}{2} (-k_p+p+2)\right)^2 \Gamma
   \left(\frac{1}{2} (k_p+p-2)\right)^4}{32 \pi ^2
   \Gamma (k_p-1) \Gamma (k_p) \Gamma (p-1) \Gamma (p)}  \,.
   }
   We can thus combine \eqref{K}, \eqref{K2}, \eqref{opeToBulk} and \eqref{beta} to get an expression finite and nonzero for each term in the sums:
  \es{MR}{
   M_R&\approx\frac{2}{\bm\Delta}\sum_{m=0} \Bigg[\sum_{k_2}\frac{k_2^2 \left(k_2^2-4\right) \Gamma \left(\frac{k_2}{2}+m-1\right) \Gamma
   \left(\frac{k_2}{2}+p\right) \Gamma \left(\frac{k_2}{2}+m-p+1\right)}{8(k_2+2 m-s)  \Gamma (m+1)
   \Gamma (p-1) \Gamma (p) \Gamma (k_2+m+1) \Gamma \left(\frac{k_2}{2}-p+1\right)}  \mathtt{t}_1  \\
  &\hspace{25mm}+\sum_{k_p}\frac{k_p (p-1) \Gamma \left(\frac{1}{2} (k_p+p+2)\right)^2 \Gamma
   \left(\frac{k_p}{2}+m-\frac{p}{2}\right)^2}{ (k_p+2 m-u) \Gamma (m+1) \Gamma (p)^2\Gamma
   (k_p+m+1) \Gamma \left(\frac{k_p-p}{2}\right)^2} \mathtt{t}_2\\
   &\hspace{25mm}+\sum_{k_p}\frac{k_p (p-1) \Gamma \left(\frac{1}{2} (k_p+p+2)\right)^2 \Gamma
   \left(\frac{k_p}{2}+m-\frac{p}{2}\right)^2}{ (k_p+2 m-t) \Gamma (m+1) \Gamma (p)^2\Gamma
   (k_p+m+1) \Gamma \left(\frac{k_p-p}{2}\right)^2} \mathtt{t}_3 \Bigg]\,,
   }
where in the crossed channels we used the crossed expressions from the $\langle2p2p\rangle$ configuration. Note that the sum over $m$ truncates to $m=0$ for $k_2=2$, as discussed above. In the block expansion of reduced correlator, this corresponds to the twist 2 term in the first line of \eqref{blockExpSQCD}, whose OPE coefficient scales with $1/c$ as in \eqref{OPEc}.\footnote{Note that the stress tensor itself does not appear in the reduced correlator, but this other protected multiplet has the same $1/c$ dependence. The $\text{dim}(G_F)$ factor in \eqref{OPEc} is cancelled by the corresponding factor in the definition of the singlet projector in Appendix A of \cite{Behan:2024vwg}.} The $\approx$ denotes the fact that these double sums are divergent, and just as in the case of the 1-loop terms in \eqref{Bs}, the divergence is just a constant that can be absorbed into the definition of the contact term $M_{F^4}$. We can regularize this divergence by checking that our final closed form answer \eqref{MR2} has the same poles in $s,t,u$ and the same growth at large $s,t,u$ as \eqref{MR}.

   We can check $\eqref{MR2}$ by comparing to the flat space Type IIB amplitude $\mathcal{A}(s,t)$ at finite $\tau$ and large $\ell_P$:
\es{Agen2}{
\mathcal{A}(s,t)&= (u+s/w)^2\Big[\ell_P^4\mathcal{A}_{F^2}+\ell_P^8\big[\cA_{F^4}(\tau)+\cA_{F^2|F^2}+\cA_R+\cA_\text{log}\log\ell_P^2+O(\ell_P^{10})\big]\Big]\,,
}
using the flat space limit formula \cite{Penedones:2010ue,Behan:2023fqq} for $\langle22pp\rangle$:
    \es{flatGen}{
\mathcal{A}^{ABCD}(s,t)=\lim_{L\to\infty} 8\pi^4 (p-1)!(p-2)!L^4\int\frac{d\beta}{2\pi i}\frac{e^\beta}{\beta^{2+p}}\frac{L^4(u+s/w)^2}{16}M^{ABCD}\Big(\frac{L^2}{2\beta}s,\frac{L^2}{2\beta}t\Big)\,,
}
which is normalized so that applying \eqref{flatGen} to \eqref{gluontree} and using the AdS/CFT dictionary \eqref{dictionary} gives
 \es{Atree}{
\mathcal{A}_{F^2}(s,t)=-\frac{(2\pi)^5}{stu}(t\,\mathtt{c}_s-s\,\mathtt{c}_t)\,.
}
The logarithmic threshold was computed in \cite{Behan:2024vwg}:
 \es{Alog}{
\mathcal{A}_{\log}(s,t)=-16\pi^6\,\bm\Delta\,(
\mathtt{t}_1+\mathtt{t}_2+\mathtt{t}_3)\,,
}
and matches \eqref{blog_fixed} after applying the flat limit to \eqref{contact} for $p=2$. To take the flat space limit of the 1-loop term written for general $p$ as a double sum in \eqref{Bs}, we note that the large $s,t$ limit gets its leading contribution from large $m,n$ in the sum, in which case the $p$-dependence is just an overall factor. We can then take large $s,t$ using the resummed expression for $p=2$ in \eqref{resummed} and simply multiply by the overall $p$-dependence, to get
\es{Aloop}{
\mathcal{A}_{F^2|F^2}(s,t)=-\frac{2\pi^6}{3}\left[\mathtt{d}_{s t} f_{\mathrm{box}}(s, t)+\mathtt{d}_{s u} f_{\mathrm{box}}(s, u)+\mathtt{d}_{t u} f_{\mathrm{box}}(t, u)\right]\,,
}
where $f_{\mathrm{box}}(s, t)$ an 8d box diagram given by taking large $s,t$ of the function $\mathcal{B}^2_{st} $ in \eqref{resummed}:
\es{f_box}{
f_{\mathrm{box}}(s, t)  \sim\mathcal{B}_{st}^2 \sim\frac{s t \log ^2\left(\frac{-s}{-t}\right)}{(s+t)^2}+2 \frac{s \log (-s/2)+t \log (-t/2)}{s+t}+\pi^2 \frac{s t}{(s+t)^2}+2\gamma\,.
}
 Note that one can arbitrarily add a contact term by shifting $\cA_{F^4}$, which we will only discuss in detail for the $D_4$ theory in section \ref{so8}.
Finally, the flat space graviton exchange term $\cA_R$ was computed in \cite{Behan:2024vwg} and takes the form 
\es{AR}{
\cA_R=-8\bm\Delta\pi^6\,\left[\mathtt{t}_1(\log (-s/2)+\gamma)+\mathtt{t}_2(\log (-t/2)+\gamma)+\mathtt{t}_3(\log (-u/2)+\gamma)\right]\,,
}
where again the choice of contact term is arbitrary.  Applying \eqref{flatGen} to \eqref{MR2} and using the AdS/CFT dictionary \eqref{dictionary} we get a precise match, which is a strong check on our result for general $p$.

\subsection{Unmixing CFT data}
\label{unmix}

The 4-point function (\ref{mellin_generalP}) contains information on the $1/N$ expansion of CFT data. Accessing this data, however, requires one to solve a mixing problem. It is well-known that interactions amongst gluon modes induces a mixing amongst double trace operators built from them. The new ingredient here is that the presence of (super-)extremal couplings induces an additional form of mixing between gluon double traces and graviton single traces. In particular, by studying the coefficients in the conformal block expansion (\ref{blockExpSQCD}), we can probe the mixing between gluon double traces that are inert under $SU(2)_L\times SU(2)_R$ and $G_F$, and the graviton modes $k_2$ in long multiplets, i.e. $k_2=4,6,\dots$.

Precisely this form of mixing due to (super-)extremal couplings has been proposed and quantified recently using a regularised bulk 3-point function computation \cite{Castro:2024cmf}. The 4-point function computation presented here independently implies this mixing, and provides an alternative (and manifestly finite) way to compute elements of the corresponding mixing matrix that matches precisely the results of \cite{Castro:2024cmf}.

For each $n=2,3,\dots$, we find that the spectrum contains $n$ long multiplets whose superprimary has dimension $\Delta=2n+\cO(1/N)$ and $\ell=0$, and is a singlet under all flavor and R-symmetries. The superprimaries of these long multiplets are a non-trivial mix of the gluon double traces schematically\footnote{See Appendix \ref{app: unmixing} for a more precise definition of these operators, which we will not need here.} of the form\footnote{We have suppressed $SU(2)_L\times SU(2)_R$ indices, which are contracted to form a singlet.}
\begin{align}
  :\!\phi^A_2 \square^{n-2}\phi^A_2\! :\,\, ,\quad  :\!\phi^A_3 \square^{n-3} \phi^A_3\! :\,\,,\,\,\quad \dots\quad \,\,, \quad  :\!\phi^A_n \phi^A_n \! : \,,
\end{align}
which we furthermore a priori allow to mix with the CFT operator $\rho_{2n}$ dual to the graviton mode $k_2 = 2n$.

We provide in Appendix \ref{app: unmixing} a full account of the interplay between these two forms of mixing, and the details of how to extract the desired unmixed CFT data for generic $n$. For clarity, let us here present our results for the simplest case of twist $4$ (i.e. $n=2$), in which there is just a single double trace operator $:\! \phi_2\phi_2 \! :$ and the graviton mode $\rho_4$ involved in the mixing. 

We then consider the $1/N$ expansion of the CFT data of the unmixed superprimaries $\cO_a$, $a=1,2$. We have scaling dimensions
\begin{align}
  \Delta_{a} = 4 + \frac{1}{N} \gamma^{(1)}_{a} + \frac{1}{N^2} \gamma^{(2)}_{a}  + \cO\left(\frac{1}{N^3}\right)\,,
\label{eq: unmixed anom dim}
\end{align}
where the index\footnote{Note that outside this section we have used the same index as a coordinate index on $S^5$, which we will have no need for here.} $a=1,2$ runs over the operator degeneracy, and we have suppressed other indices that specify spin, flavor representation etc. . We also expand their superblock OPE coefficients appearing in the expansion (\ref{blockExpSQCD}) as
\begin{align}
  \lambda_{2,2,\cO_a}=\lambda^{(0)}_{2,2,\cO_a} + \frac{1}{N}\lambda^{(1)}_{2,2,\cO_a} + \frac{1}{N^2}\lambda^{(2)}_{2,2,\cO_a} +  \cO\left(\frac{1}{N^3}\right) \,,
\label{eq: unmixed OPE coeff}
\end{align}
again suppressing other quantum number labels. For brevity, let us write $\lambda^{(0)}_a = \lambda^{(0)}_{2,2,\cO_a}$ and similarly for the higher corrections.

To access these data, we expand the reduced correlator of $\langle 2222 \rangle $ in its flavor singlet channel into conformal blocks as in (\ref{blockExpSQCD}), and furthermore expand in small $U$, to find
\begin{align}
  &\cG^{2,2}_\mathbf{1}(U,V)= U^{-1} \Bigg[\left\langle (\lambda^{(0)})^2  \right\rangle  + \frac{1}{N}\left(2\left\langle \lambda^{(0)}\lambda^{(1)}  \right\rangle + \frac{1}{2}\left\langle (\lambda^{(0)})^2\gamma^{(1)} \right\rangle  (\log U + 2\partial^\text{no log}_\Delta)\right) \nn\\
  &\hspace{30mm} + \frac{1}{N^2} \bigg( \left\langle 2\lambda^{(0)}\lambda^{(2)}  + (\lambda^{(1)})^2 \right\rangle	+ \frac{1}{2}\left\langle 2\lambda^{(0)}\lambda^{(1)}\gamma^{(1)} + (\lambda^{(0)})^2\gamma^{(2)}  \right\rangle(\log U + 2\partial^\text{no log}_\Delta)			\nn\\
  &\hspace{45mm} + \frac{1}{8} \left\langle (\lambda^{(0)})^2(\gamma^{(1)})^2 \right\rangle	(\log^2 U + 4\log U \partial_\Delta + 4 (\partial^\text{no log}_\Delta)^2) \bigg)			\nn\\
   &\hspace{30mm} +  \cO\!\left(\frac{1}{N^3}\right)\Bigg]g_{\Delta+2,0}(U,V) \bigg|_{\Delta=4}				\nn\\
   &\hspace{20mm} + \dots  \, ,
\label{eq: U expansion}
\end{align}
where the ellipses contain contributions from short multiplets, as well as all long multiplets of differing spin and/or engineering twist. Here, $\partial^\text{no log}_\Delta$ means that we take the derivative, and then throw away terms involving $\log U$, as these are already accounted for in other terms.

In (\ref{eq: U expansion}), the $\cO(1)$ term is computed in the generalized free field theory; the $\cO(1/N)$ term comes from gluon exchange; and the $\cO(1/N^2)$ term receives contributions from both 1-loop gluon processes\footnote{These contributions are fixed entirely in terms of data from tree level gluon exchange.} and graviton exchange. In this expression, the angled brackets denote averages over the $2$-fold multiplicity of data, so that for instance
\begin{align}
  \left\langle (\lambda^{(0)})^2\right\rangle = \sum_{a=1}^2 (\lambda^{(0)}_a)^2\,.
\end{align}
We then have enough data to extract the leading CFT data $\gamma_a^{(1)}$ and $\lambda_a^{(0)}$. The expansion coefficients in which solely these data appear are found to be
\begin{align}
  \left\langle (\lambda^{(0)})^2  \right\rangle 					 &= \frac{2}{3}		\nn\,,\\
  \left\langle (\lambda^{(0)})^2 \gamma^{(1)}  \right\rangle 	 &= -\frac{4}{3}\frac{h^\vee}{\mathbf{\Delta}} =  \frac{8(1-\mathbf{\Delta})}{\mathbf{\Delta}}				\nn\,,\\
  \left\langle (\lambda^{(0)})^2 (\gamma^{(1)})^2  \right\rangle &= \left\langle (\lambda^{(0)})^2 (\gamma^{(1)})^2  \right\rangle_\text{gluon} + \left\langle (\lambda^{(0)})^2 (\gamma^{(1)})^2  \right\rangle_\text{grav}\,,
\end{align}
where in the final expression we have separated the contributions from 1-loop gluon and graviton exchange, which are given by
\begin{align}
  \left\langle (\lambda^{(0)})^2 (\gamma^{(1)})^2  \right\rangle_\text{gluon} &= \frac{3}{2}\left(-\frac{4}{3}\frac{h^\vee}{\mathbf{\Delta}}\right)^2 = \frac{96(1-\mathbf{\Delta})^2}{\mathbf{\Delta}^2}	\,,	\nn\\
  \left\langle (\lambda^{(0)})^2 (\gamma^{(1)})^2  \right\rangle_\text{grav} &= \frac{16}{5} \frac{\dim(G_F)}{\mathbf{\Delta}} = \frac{32}{5}\frac{(6-5\mathbf{\Delta})(5-6\mathbf{\Delta})}{\mathbf{\Delta}^2}\,.
\end{align}
 To arrive at these final expressions, we have used
\begin{align}
  h^\vee = 6(\mathbf{\Delta}-1),\qquad \dim(G_F) = \frac{2(6-5\mathbf{\Delta})(5-6\mathbf{\Delta})}{\mathbf{\Delta}}\, .
\end{align}
We can then read off from these data the corresponding mixing matrix $M_{ab}$, defined by\footnote{Here, the normalisation factor $\cN^2 = 2\times 3 \times \dim(G_F)$ ensures that $\frac{1}{\cN}:\!\phi_2  \phi_2 \!:$ has unit 2-point function at leading order.}
\begin{align}
  &\begin{pmatrix}
  	\langle \rho_{4}(x) \rho_{4}(y) \rangle 	& \frac{1}{\cN}\langle \rho_{4}(x) :\!\phi_2  \phi_2 \!:\!(y) \rangle	\\
  	\frac{1}{\cN}\langle :\!\phi_2  \phi_2 \!:\!(x)\, \rho_{4}(y)  \rangle & \frac{1}{\cN^2}\langle :\!\phi_2   \phi_2 \!:\!(x) :\!\phi_2 \phi_2 \!:\!(y) \rangle		
  \end{pmatrix}\nn\\
  &\qquad = \frac{1}{|x-y|^{8}} \left[\, \delta_{ab} - \frac{1}{N} M_{ab} \log\left(|x-y|^2\right) + \cO\left(\frac{1}{N^2}\right) \right]\,.
\end{align}
It is given by
\begin{align}
  M &= \frac{1}{\left\langle (\lambda^{(0)})^2  \right\rangle} \begin{pmatrix}
  	0			& 		\sqrt{\left\langle (\lambda^{(0)})^2 (\gamma^{(1)})^2  \right\rangle_\text{grav}\left\langle (\lambda^{(0)})^2  \right\rangle}		\nn\\
  	\sqrt{\left\langle (\lambda^{(0)})^2 (\gamma^{(1)})^2  \right\rangle_\text{grav}\left\langle (\lambda^{(0)})^2  \right\rangle}	& \left\langle (\lambda^{(0)})^2 \gamma^{(1)}  \right\rangle
  \end{pmatrix} \nn\\
  &= \begin{pmatrix}
  	0		&		\frac{4}{\mathbf{\Delta}}\sqrt{\frac{3}{5}}\sqrt{(6-5\mathbf{\Delta})(5-6\mathbf{\Delta})}		\\
  	\frac{4}{\mathbf{\Delta}}\sqrt{\frac{3}{5}}\sqrt{(6-5\mathbf{\Delta})(5-6\mathbf{\Delta})}		& 		\frac{12(1-\mathbf{\Delta})}{\mathbf{\Delta}}
  \end{pmatrix}\,.
  \label{eq: twist 4 mixing matrix}
\end{align}
The zero in the top left corner arises precisely because the leading correction to the 2-point function $\langle \rho_{4}\rho_{4} \rangle $ comes from 1-loop diagrams (both gluon and graviton), which begin at order $1/N^2$.

We have checked that the off-diagonal entry in the mixing matrix, which parameterises the mixing due to the extremal coupling, precisely matches an independent bulk computation. In detail, it is reproduced precisely when one plugs the value of our bulk coupling into the holographically renormalised result\footnote{One also has to rescale this result, since we are using a convention with unit-normalised 2-point functions.} (3.25) of \cite{Castro:2024cmf}. Since our 4-point exchange diagrams were manifestly finite, this provides a rather non-trivial check of the validity of the regularisation used in \cite{Castro:2024cmf}.

Getting back to our unmixing problem, the leading anomalous dimensions $\gamma^{(1)}_a$ are the eigenvalues of $M$, while the leading OPE coefficients $\lambda^{(0)}_a$ are encoded in the corresponding eigenvectors according to (\ref{eq: unmixed data answer}). We find ultimately
\begin{align}
  \gamma^{(1)}_{1} &=  \frac{6}{\mathbf{\Delta}}\left(1-\mathbf{\Delta} -  \frac{1}{\sqrt{15}}\sqrt{135 - 274\mathbf{\Delta} + 135\mathbf{\Delta}^2}\right),\nn\\
   \gamma^{(1)}_{2} &= \frac{6}{\mathbf{\Delta}}\left(1-\mathbf{\Delta} +  \frac{1}{\sqrt{15}}\sqrt{135 - 274\mathbf{\Delta} + 135\mathbf{\Delta}^2}\right)\,,
\label{eq: twist 4 anom dims}
\end{align}
and
\begin{align}
  (\lambda^{(0)}_{2,2,\cO_1})^2 				&= \frac{2}{3}\left( \frac{-\sqrt{15}(1-\mathbf{\Delta})+\sqrt{135-274\mathbf{\Delta}+135\mathbf{\Delta}^2}}{2\sqrt{135-274\mathbf{\Delta}+135\mathbf{\Delta}^2}}\right),\nn\\
   \qquad (\lambda^{(0)}_{2,2,\cO_2})^2 		&= \frac{2}{3}\left( \frac{\sqrt{15}(1-\mathbf{\Delta})+\sqrt{135-274\mathbf{\Delta}+135\mathbf{\Delta}^2}}{2\sqrt{135-274\mathbf{\Delta}+135\mathbf{\Delta}^2}} \right)\,.
\label{eq: twist 4 OPE coeff}
\end{align}
Let us finally say a few words about the case of generic $n$, i.e. generic engineering twist $2n$. We would like to once again fully unmix at leading order; that is, determine the leading superblock OPE coefficients $\lambda^{(0)}_{p,p, \cO_a}$ for all $p=2,\dots,n$ and leading anomalous dimensions $\gamma^{(1)}_a$ of the $n$ unmixed superprimaries $\cO_a$. To do so, one needs precisely the following data:
\begin{itemize}
  \item The  $\cO(1)$ part of $\langle pppp \rangle $ for $p=2,\dots,n$, which can be computed in generalized free field theory.
  \item The coefficient of $\frac{1}{N}(\log U)\, g_{2n+2,0}(U,V)$ in the expansion of $\langle ppqq\rangle$ for all $p,q=2,\dots,n$. This comes from gluon exchange. These numbers fix the mixing coefficients in $M$ amongst the double traces.
  \item The contribution from graviton exchange to the coefficient of $\frac{1}{N^2}(\log^2 U) \,g_{2n+2,0}(U,V)$ in the expansion of $\langle ppqq\rangle$. These numbers fix the mixing coefficients in $M$ between $\rho_{2n}$ and the double traces. It is in fact overkill to compute these coefficients for all $p,q=2,\dots,n$, as they are not all independent. In particular, it is sufficient to just consider $\langle 22pp\rangle$ for all $p=2,\dots,n$.
\end{itemize}
For all $n$, the first two pieces of data are already known in \cite{Alday:2021ajh,Huang:2023ppy,Drummond:2022dxd}. We have then computed here the graviton exchange contribution to $\langle 22pp\rangle $ for all $p=2,\dots$, and thus in principle have sufficient data to unmix the data $\lambda^{(0)}_a$ and $\gamma^{(1)}_a$ for generic $n$. 

We do not seek here generic closed formulae for these data as a function of $n$. But as a final example, we find that at twist $6$ (i.e. $n=3$), the unmixed superprimaries $\cO_a$, $a=1,2,3$, have leading CFT data given in terms of the eigenvalues and eigenvectors of the mixing matrix $M_3$ given in (\ref{eq: M3}), according to the formulae (\ref{eq: unmixed data answer}). For concreteness, in the case of $G_F = D_4$ (i.e. $\mathbf{\Delta}=2$) we find
\begin{align}
  \gamma^{(1)}_1 = -106.132\dots \,\, , \qquad \gamma^{(1)}_2 = 72.274\dots  \,\, , \qquad \gamma^{(1)}_3 = -4.542\dots\,,
\end{align}
and
\begin{align}
   (\lambda^{(0)}_{2,2,\cO_1})^2		= 0.0086\dots \,\,,\qquad  (\lambda^{(0)}_{2,2,\cO_2})^2	= 0.0016\dots \,\,,\qquad (\lambda^{(0)}_{2,2,\cO_3})^2			= 0.0898\dots \,,
\end{align}
\begin{align}
   (\lambda^{(0)}_{3,3,\cO_1})^2		= 0.1283\dots \,\,,\qquad  (\lambda^{(0)}_{3,3,\cO_2})^2	= 0.0985\dots \,\,,\qquad (\lambda^{(0)}_{3,3,\cO_3})^2			=  0.0232\dots \,.
\end{align}

\subsection{Fixing the $D_4$ theory to $O(1/N^2)$}
\label{so8}

We finish this section by specializing to the $D_4$ theory for $p=2$, where we can use supersymmetric localization to completely fix $M_{F^4}$, generalizing \cite{Behan:2023fqq} that only considered the $\tau$ dependence. The irreps that appear in the correlator for this theory are
\es{tensso8}{
{\bf28}\otimes{\bf28}={\bf1}\oplus {\bf28}\oplus {\bf35_v}\oplus {\bf35_c}\oplus {\bf35_s}\oplus {\bf300}\oplus {\bf350}\,.
}
The corresponding projectors are
\begin{align}\label{Pbas}
\begin{split}
P_{\mathbf{1}} & =\left(\begin{array}{lllllll}
\frac{1}{28} & 0 & 0 & 0 & 0 & 0 & 0
\end{array}\right) \,\cdot \mathtt{t}\,,\\
P_{\mathbf{2 8}} & =\left(\begin{array}{lllllll}
0 & 0 & 0 & -\frac{1}{12} & \frac{1}{12} & 0 & 0
\end{array}\right) \cdot \mathtt{t}\,,\\
P_{\mathbf{3 5}_ \mathbf{c}} & =\left(\begin{array}{lllllll}
0 & \frac{1}{12} & \frac{1}{12} & 0 & \frac{1}{12} & -\frac{1}{12} & \frac{1}{2}
\end{array}\right) \cdot \mathtt{t}\,,\\
P_{\mathbf{3 5}_{\mathbf{s}}} & =\left(\begin{array}{lllllll}
0 & \frac{1}{12} & \frac{1}{12} & 0 & \frac{1}{12} & -\frac{1}{12} & -\frac{1}{2}
\end{array}\right)\cdot \mathtt{t}\,, \\
P_{\mathbf{3 5}_ \mathbf{v}} & =\left(\begin{array}{lllllll}
-\frac{1}{12} & 0 & 0 & \frac{1}{12} & -\frac{1}{12} & \frac{1}{6} & 0
\end{array}\right)\cdot \mathtt{t}\,, \\
P_{\mathbf{3 5 0}} & =\left(\begin{array}{lllllll}
0 & -\frac{1}{2} & \frac{1}{2} & \frac{1}{12} & -\frac{1}{12} & 0 & 0
\end{array}\right) \cdot \mathtt{t}\,,\\
P_{\mathbf{3 0 0}} & =\left(\begin{array}{lllllll}
\frac{1}{21} & \frac{1}{3} & \frac{1}{3} & -\frac{1}{12} & -\frac{1}{12} & 0 & 0
\end{array}\right)\cdot \mathtt{t}\,,
\end{split}
\end{align}
where the tensor structures are
\begin{align}\label{tbas}
\begin{split}
\mathtt{t}_i=&\left(\delta^{AB}\delta^{CD},\quad \delta^{AC}\delta^{BD},\quad \delta^{AD}\delta^{BC},\quad 
f^{ACE}f^{BDE},\quad f^{ADE}f^{BCE},\right.\\
&\left. \text{tr}(T^AT^BT^CT^D),\quad \frac{1}{4\cdot 4!}\epsilon_{a_1a_2b_1b_2c_1c_2d_1d_2} T^A_{a_1a_2}T^B_{b_1b_2}T^C_{c_1c_2}T^D_{d_1d_2}\right)\,.
\end{split}
\end{align}
The contact Mellin amplitudes in this basis include $M_{F^4}^1$ defined in \eqref{p2log}, as well as 
\es{treeEx}{
 M^2_{F^4}=\mathtt{t}_6+\frac13\mathtt{t}_4-\frac23\mathtt{t}_5\,,\quad M^3_{F^4}=\mathtt{t}_7\,. 
 }
We can relate derivatives of the mass deformed sphere free energy $F(\mu_i)$ to certain integrals of the Mellin space correlator in the basis \eqref{tbas} as \cite{Chester:2022sqb}
 \es{intNew}{
-\partial_{\mu_1}^4 F\big|_{\mu=0}&=k^2I[M_1+M_2+M_3+2M_6]\,,\\
  -\partial_{\mu_1}^2\partial_{\mu_2}^2 F\big|_{\mu=0}&=\frac{k^2}{3}I[M_1+M_2+M_3]\,,\\
-\partial_{\mu_1}\partial_{\mu_2}\partial_{\mu_3}\partial_{\mu_4} F \big|_{\mu=0}&=\frac{k^2}{6}I[M_7]\,,\\
}
where the integral is
\es{Imel}{
  I[M_{\bf r}]\equiv -\int \frac{ds dt}{(4\pi i)^2}& \Bigg[M_{\bf r}(s,t) \Gamma[2-s/2]\Gamma[s/2]\Gamma[2-t/2]\Gamma[t/2]\Gamma[2-{u}/2]\Gamma[{u}/2] \\
 & \times \Big(\frac{H_{\frac s2-1}+H_{1-\frac s2}}{(t-2)(u-2)}+\frac{H_{\frac t2-1}+H_{1-\frac t2}}{(s-2)(u-2)}+\frac{H_{\frac u2-1}+H_{1-\frac u2}}{(s-2)(t-2)}\Big)\Bigg]\,.
}
The integrated constraint can also be written in the projector basis \eqref{Pbas} as
\es{Fcsv}{
\mathcal{F}_{\bf v}&\equiv   -4\partial_{\mu_1}^2\partial_{\mu_2}^2 F\big|_{\mu=0}\,,\\
 \mathcal{F}_{\bf c}&\equiv    -\partial_{\mu_1}^4 F\big|_{\mu=0}- \partial_{\mu_1}^2\partial_{\mu_2}^2 F\big|_{\mu=0}+2  \partial_{\mu_1}\partial_{\mu_2}\partial_{\mu_3}\partial_{\mu_4} F \big|_{\mu=0}\,,\\
  \mathcal{F}_{\bf s}&\equiv    -\partial_{\mu_1}^4 F\big|_{\mu=0}- \partial_{\mu_1}^2\partial_{\mu_2}^2 F\big|_{\mu=0}-2  \partial_{\mu_1}\partial_{\mu_2}\partial_{\mu_3}\partial_{\mu_4} F \big|_{\mu=0}\,,\\
}
which is naturally permuted by $SL(2,\mathbb{Z})$ transformations of $\tau$ as \cite{Seiberg:1994aj,Sen:1996vd}
\es{dualTrial}{
&S:\qquad\tau\to-1/\tau\qquad \Leftrightarrow \qquad{\bf 35_v} \leftrightarrow {\bf 35_c} \qquad\Leftrightarrow   \qquad{\mathcal{F}_{\bf v}} \leftrightarrow {\mathcal{F}_{\bf c}}\,,\\
&T:\qquad\tau\to\tau+1\qquad \Leftrightarrow \qquad{\bf 35_c} \leftrightarrow {\bf 35_s}\qquad\Leftrightarrow   \qquad{\mathcal{F}_{\bf c}} \leftrightarrow {\mathcal{F}_{\bf s}}\,. \\
}
The RHS was computed using supersymmetric localization in \cite{Beccaria:2022kxy,Behan:2023fqq}:
\es{Ffinal}{
\mathcal{F}_{\bf v}&=-8\log[\tau_2 |\theta_3(\tau)\theta_4(\tau)|^2]+8\log[8\pi N]+4f(N)+O(e^{-N})\,,\\
\mathcal{F}_{\bf c}&=-8\log[\tau_2 |\theta_2(\tau)\theta_3(\tau)|^2]+8\log[8\pi N]+4f(N)+O(e^{-N})\,,\\
\mathcal{F}_{\bf s}&=-8\log[\tau_2 |\theta_2(\tau)\theta_4(\tau)|^2]+8\log[8\pi N]+4f(N)+O(e^{-N})\,,\\
}
where we define
\es{tF}{
f(N)&=\frac{16}{3}-4 \zeta (3)+4 \gamma -4 \log (4 \pi )-2\log(2N)+2\psi(2 N+5/2)\\
&+(N+1/4)\psi^{(1)}(N+{5}/{4})+(N+3/4)\psi^{(1)}(N+{7}/{4})\\
&= \frac{22}{3}  -4 \zeta (3)+4 \gamma -4 \log (4 \pi )+ \frac{1}{N} - \frac{7}{48N^2} - \frac{1}{48N^3} + \frac{67}{2560N^4} + O(N^{-5})\,.
}
The integral acting on the Mellin amplitudes that appear in \eqref{mellin_generalP} for $p=2$ gives
\es{Is}{
I[1]=\frac{1}{24}\,,\qquad I[\mathcal{B}(s,t)]=\frac{11}{1296}-\frac{\zeta (3)}{108}\,\qquad I[3H_{1-\frac s2}+\frac{4}{s-2}]=-\frac{11}{96}\,,
}
and similar for the crossed functions of $t,u$. We thus fix the $b_{F^4}^i$ coefficients in \eqref{mellin_generalP} for $p=2$ to 
\es{bvStrong}{
b^1_{F^4}&=6 (\gamma - \log (4 \pi ))-3\log\Big[\tau_2 |\theta_3(\tau)\theta_4(\tau)|^2\Big]\,,\\
 b^2_{F^4}&=-3\log\Big[\tau_2 |\theta_2(\tau)\theta_3(\tau)|^2\Big]-3\log\Big[\tau_2|\theta_2(\tau)\theta_4(\tau)|^2\Big]+6\log\Big[\tau_2 |\theta_3(\tau)\theta_4(\tau)|^2\Big]\,,\\
  b^3_{F^4}&=18\log\Big[\tau_2 |\theta_2(\tau)\theta_3(\tau)|^2\Big]-18\log\Big[\tau_2 |\theta_2(\tau)\theta_4(\tau)|^2\Big]\,,\\
}
where the $\tau$ dependence was previously fixed in \cite{Behan:2023fqq}, and note that the $\zeta(3)$ cancelled between \eqref{tF} and \eqref{Is}.

We can now take the flat space limit using \eqref{flatGen} for $p=2$ to fix $\cA_{F^4}$ in \eqref{Agen2}. Note that when applying the flat limit formula to the $\log s$ terms that appear in $M_{R}$ and $M_{F^2|F^2}$, we should keep in mind that 
 \es{logInt}{
 \int\frac{d\beta}{2\pi i}\frac{e^\beta}{\beta^4}\log(1/(2\beta))=-\frac{11}{36}+\frac{\gamma }{6}-\frac{\log (2)}{6}\,.
 }
 With the choice of contact term in $A_R$ \eqref{AR} and $A_{F^2|F^2}$ \eqref{Aloop}, we find that $\cA_{F^4}$ is
 \es{AF4}{
 \cA_{F^4}=a^1_{F^4}(\mathtt{t}_1+\mathtt{t}_2+\mathtt{t}_3)+a^2_{F^4}\Big(\mathtt{t}_6+\frac13\mathtt{t}_4-\frac23\mathtt{t}_5\Big)+a^1_{F^4}\mathtt{t}_7\,,
 }
 with coefficients
 \es{avStrong}{
a^1_{F^4}&=-\frac{16}{3}\pi^6(6\log(2\pi)-11)-16\pi^6\log\Big[\tau_2 |\theta_3(\tau)\theta_4(\tau)|^2\Big]\,,\\
 a^2_{F^4}&=-16\pi^6\log\Big[\tau_2 |\theta_2(\tau)\theta_3(\tau)|^2\Big]-16\pi^6\log\Big[\tau_2|\theta_2(\tau)\theta_4(\tau)|^2\Big]+32\pi^6\log\Big[\tau_2 |\theta_3(\tau)\theta_4(\tau)|^2\Big]\,,\\
  a^3_{F^4}&=96\pi^6\log\Big[\tau_2 |\theta_2(\tau)\theta_3(\tau)|^2\Big]-96\pi^6\log\Big[\tau_2 |\theta_2(\tau)\theta_4(\tau)|^2\Big]\,.\\
}
Note that all the $\gamma$ terms naturally appear next to $\log s$ terms in \eqref{AR} and \eqref{Aloop}. The $\tau$ dependent part of these factors exactly matches the independent result in \cite{Kiritsis:2000zi}, as was first checked in \cite{Behan:2023fqq}. The $\tau$-independent terms are a new prediction though.\footnote{Note that \cite{Kiritsis:2000zi} did not carefully consider the $\tau$-independent terms. We thank Boris Pioline for discussions about this.}

Now that we have completely fixed the correlator to order $1/N^2$, we can read off CFT data of twist four long multiplets.\footnote{Higher twist experience degeneracy and so we will not discuss them.} We can use the projection method of \cite{Chester:2018lbz} to extract the long multiplet anomalous dimensions $\gamma^{(2)}_{r,\ell}$ at order $1/N^2$ from the amplitudes $M_{F^2|F^2}$, $M_R$, and $M_{F^4}$. For $\ell>0$ only $M_{F^2|F^2}$ and $M_R$ contribute and we get for the even spin irreps
\es{anomN2}{
\gamma_{\textbf{1},\ell}^{(2)} &=-\frac{8 \left(\ell^6+15 \ell^5+56 \ell^4+7 \ell^3+357 \ell^2+2360
   \ell+1536\right)}{\ell (\ell+1)^3 (\ell+4)^3 (\ell+5)}\,,\\
\gamma^{(2)}_{\textbf{35}_*,\ell} &=-\frac{8 \left(\ell^6+15 \ell^5+82 \ell^4+203 \ell^3+295 \ell^2+400 \ell+192\right)}{\ell (\ell+1)^3 (\ell+4)^3
   (\ell+5)}\,, \\
\gamma^{(2)}_{\textbf{300},\ell} &=-\frac{8 \left(\ell^6+15 \ell^5+91 \ell^4+287 \ell^3+511 \ell^2+505 \ell+192\right)}{\ell (\ell+1)^3 (\ell+4)^3
   (\ell+5)} \,, \\
}
and for the odd spin irreps
\es{anomN22}{
\gamma^{(2)}_{\textbf{28},\ell} &= -\frac{8 \left(\ell^6+15 \ell^5+77 \ell^4+163 \ell^3+273 \ell^2+665 \ell+384\right)}{\ell (\ell+1)^3 (\ell+4)^3
   (\ell+5)}\,, \\
\gamma^{(2)}_{\textbf{350},\ell} &=-\frac{8 (\ell+2) (\ell+3)}{\ell (\ell+1) (\ell+4) (\ell+5)}\,.
}
For zero spin recall that the singlet irrep data is ambiguous due to mixing, but for the other irreps we unambiguously get
\es{anomzero}{
\gamma^{(2)}_{\textbf{300},0}&=\frac{12}{5} (2 \log [\sqrt{\tau_2}|\eta(\tau)|^2 ]-2 \gamma +2 \log \pi +5\log 2)+\frac{723}{200}\,,\\
\gamma^{(2)}_{\textbf{35}_\textbf{v},0}&=-\frac{6}{5} (8\log [\sqrt{\tau_2}|\eta(\tau)|^2 ]-12\log [\sqrt{\tau_2}|\theta_2(\tau)|^2 ]+4 \gamma -4 \log \pi +2\log
   2)+\frac{126}{25}\,,\\
   \gamma^{(2)}_{\textbf{35}_\textbf{c},0}&=-\frac{6}{5} (8\log [\sqrt{\tau_2}|\eta(\tau)|^2 ]-12\log [\sqrt{\tau_2}|\theta_4(\tau)|^2 ]+4 \gamma -4 \log \pi +2\log
   2)+\frac{126}{25}\,,\\
   \gamma^{(2)}_{\textbf{35}_\textbf{s},0}&=-\frac{6}{5} (8\log [\sqrt{\tau_2}|\eta(\tau)|^2 ]-12\log [\sqrt{\tau_2}|\theta_3(\tau)|^2 ]+4 \gamma -4 \log \pi +2\log
   2)+\frac{126}{25}\,,\\
}
where the three 35-dimensional irreps are permuted under $SL(2,\mathbb{Z})$ as \eqref{dualTrial}.

\section{Conclusion}\label{sec:conclusion}

In this paper we computed the extremal and super-extremal bulk couplings $\beta_{pqk_r}$ between graviton and gluon modes in various F-theory AdS/CFT duals. We then used these bulk couplings to compute the graviton exchange contribution to gluon scattering $\langle 22pp\rangle $ in $AdS_5\times S^3$, and found the flat space limit of our answer matched the prediction from \cite{Behan:2024vwg}. We also unmixed the lowest single trace graviton modes and double trace gluon modes, and computed the leading $1/N$ anomalous dimensions of each. For $\langle 2222\rangle $ in the $D_4$ theory, we furthermore used localization to completely fix the correlator at $1/N^2$ order, and used this to read off CFT data at this order and give a prediction for the flat space limit.

Looking ahead, it would be interesting to generalize our calculation to the most general $\langle ppqq\rangle$. The 1-loop gluon term, which appears at the same order in $1/N^2$, was computed for certain values of $p$ and $q$ in \cite{Huang:2023ppy}, but not yet in general. We would also like to unmix the single trace graviton modes and the double trace gluon modes to $1/N^2$ order. This will require additional data, such as the $1/N^2$ correction to the graviton mode mass, which must be computed from a 1-loop calculation in the bulk. In general, this project demonstrates the limit of the bootstrap approach to holographic correlators, as the bulk couplings $\beta_{ppk_2}$ and $\beta_{2pk_p}$ needed to compute the graviton exchange diagram could not be fixed from any CFT principle, but required an independent bulk calculation. It would also be interesting to see how the $1/N^2$ data considered in this work affects the calculation of the 2-loop gluon exchange term at order $1/N^3$, which was initiated in \cite{Huang:2023oxf}. The effect of graviton modes on unmixing should also effect the putative approximate 6d conformal symmetry considered in \cite{Drummond:2022dxd}.

The phenomena of non-vanishing extremal couplings and an infinite tower of graviton exchanges is quite general. For instance, we expect it to apply to gluons scattering on $AdS_{d+1}\times S^3$ dual to CFT$_d$ for $d=3,4,5,6$. In $d=5,6$, the graviton exchange term is the first correction to the tree gluon exchange term, and there are no contact terms at this order, so we should be able to unambiguously compute all CFT data at this order. In $d=3$, the graviton exchange term occurs at the same order as the 2-loop gluon exchange, as first discussed in \cite{Chester:2023qwo}. Since the graviton exchange terms are sensitive to the compact dimensions in the bulk dual, it would be interesting to compare their corrections to the CFT data to non-perturbative bounds from the numerical bootstrap, similar to what was done for graviton scattering in maximally supersymmetric theories in \cite{Alday:2022ldo}.

Finally, it would be interesting to find other examples of AdS/CFT duals with even less supersymmetry that have extremal couplings, and see how these affect correlation functions.

\section*{Acknowledgments} 
We thank Ofer Aharony, Fernando Alday, Connor Behan, Nikolay Bobev, Alejandra Castro, Pietro Ferrero, Daniele Pavarini, Hynek Paul, Boris Pioline, Silviu Pufu, and David Turton for useful discussions. SMC is supported by the Royal Society under the grant URF\textbackslash R1\textbackslash 221310 and the UK Engineering and Physical Sciences Research council grant number EP/Z000106/1, RM is also supported by EP/Z000106/1, JvM is supported by the STFC Consolidated Grant ST/X000575/1, and acknowledges the ERC-COG grant NP-QFT No. 864583 for its support in the early stages of this project.  JvM would like to thank the ITF at the KUL for their continuous hospitality.

\appendix \addtocontents{toc}{\protect\setcounter{tocdepth}{1}}

\section{Details of the computation of cubic couplings }\label{app: cubic couplings}

In this appendix we provide the full details of the computation of the bulk cubic couplings (\ref{eq: coupling final answer}).

\subsection{Geometric setup}

The basic task at hand boils down to computing overlaps of various scalar and vector harmonics, defined both on the 5-sphere and an embedded 3-sphere inside it. For this task, we find it useful to work in an ambient $\R^6$ with Cartesian coordinates $y_I$, $a=1,\dots 6$. The unit $S^5$ is defined by the constraint
\begin{align}
  y_1^2 + y_2^2 + y_3^2 +y_4^2 = 1-\rho^2,\qquad y_5^2 +y_6^2 = \rho^2
\end{align}
where the coordinate $\rho$ goes over the range $\rho\in [0,1]$. In this way, we realise $S^5$ as a (singular) $S^3\times S^1$ fibration over an interval, where the $S^3$ degenerates at one end of the interval, and the $S^1$ at the other.

We are actually interested in an orbifold $\tilde{S}^5$ of $S^5$, which amounts to changing the periodicty of the polar coordinate in the $(y_5,y_6)$ plane from $2\pi$ to $2\pi /\mathbf{\Delta}$. It is clear then that the fixed point locus of this orbifold action is the surface $\rho=0$, which is just a unit $S^3$. The orbifold breaks the isometries as
\begin{align}
  SO(6) \cong SU(4) \longrightarrow \quad SO(4) \times SO(2) \cong SU(2)_L \times SU(2)_R \times U(1)
\end{align}
The virtue of this ambient space formulation is that there is just one integral we will ever need. Namely, if $y_I$ are Cartesian coordinates in $\R^{D+1}$ and $S^D$ is defined by $x_I x_I = 1$, then
\begin{align}
  &\int_{S^D} d\Omega_D\,y_{I_1}y_{I_2}\dots y_{I_{2n}} \nn\\
  &\quad = \frac{\text{Vol}(S^D)}{(D+1)(D+3)\dots (D+2n-1)}\left(\delta_{I_1 I_2}\dots \delta_{I_{2n-1}I_{2n}} + \text{all other contractions}\right)
\label{eq: general integral identity}
\end{align}
Let us present some special cases of this formula that will be useful later. Let $i=1,2,3,4$. The coordinates $y_i$ then transform in the vector representation of $SO(4)\subset SO(6)$. In what follows, we will express representations of $SO(4)$ manifestly as symmetric tensors of $SU(2)_L$ and $SU(2)_R$, where $\frak{so}(4) \cong \frak{su}(2)_L \oplus \frak{su}(2)_R$. To this end, it is useful to define
\begin{align}
  y^{\bar{\alpha}\alpha} = (\sigma_i)^{\bar{\alpha}\alpha} y_i
\end{align}
with $\bar{\alpha},\bar{\beta},\dots$ fundamental indices of $SU(2)_L$, and $\alpha,\beta,\dots $ fundamental indices of $SU(2)_R$. Here, $\sigma_i$ is the $SO(4)$ (i.e. 4-dimensional Euclidean) Pauli 4-vector, $(\sigma_i)^{\bar{\alpha}\alpha} = (i\sigma^P_1,i\sigma^P_2,i\sigma^P_3,\mathds{1}_2)$ in terms of the regular Pauli matrices $\sigma^P_a$. All we'll really need to know about these is that
\begin{align}
  (\sigma_i)^{\bar{\alpha} \alpha}(\sigma_i)^{\bar{\beta}\beta} = 2 \epsilon^{\bar{\alpha}\bar{\beta}} \epsilon^{\alpha\beta}
\end{align}
Then, on $\tilde{S}^5$, (\ref{eq: general integral identity}) implies that for any function $f(\rho^2)$ we have
\begin{align}
  &\int_{\tilde{S}^5} d\Omega_5\, y^{\bar{\alpha}_1 \alpha_1 }\dots y^{\bar{\alpha}_{2n} \alpha_{2n}} f(\rho^2) \nn\\
  & = \frac{2  \pi^3}{(n+1)!\mathbf{\Delta}}\left(\int_0^{1} dt\, (1-t)^{n+1}f(t)\right)\left(\epsilon^{\bar{\alpha}_1\bar{\alpha}_2} \epsilon^{\alpha_1\alpha_2}\dots \epsilon^{\bar{\alpha}_{2n-1}\bar{\alpha}_{2n}} \epsilon^{\alpha_{2n-1}\alpha_{2n}}  + \text{all other contractions}\right)  
\label{eq: S5 gen int SU(2)}
\end{align}
where for instance for the case $n=2$, the final part in brackets reads in full
\begin{align}
  \epsilon^{\bar{\alpha}_1\bar{\alpha}_2} \epsilon^{\alpha_1\alpha_2}\epsilon^{\bar{\alpha}_3\bar{\alpha}_4} \epsilon^{\alpha_3\alpha_4} + \epsilon^{\bar{\alpha}_1\bar{\alpha}_3} \epsilon^{\alpha_1\alpha_3}\epsilon^{\bar{\alpha}_2\bar{\alpha}_4} \epsilon^{\alpha_2\alpha_4}+\epsilon^{\bar{\alpha}_1\bar{\alpha}_4} \epsilon^{\alpha_1\alpha_4}\epsilon^{\bar{\alpha}_2\bar{\alpha}_3} \epsilon^{\alpha_2\alpha_3}
\end{align}
Secondly, on the fixed point locus $S^3 \subset \tilde{S}^5$ we have
\begin{align}
  \int_{S^3}d\Omega_3\,  y^{\bar{\alpha}_1 \alpha_1 }\dots y^{\bar{\alpha}_{2n} \alpha_{2n}} = \frac{2\pi^2}{(n+1)!}\left(\epsilon^{\bar{\alpha}_1\bar{\alpha}_2} \epsilon^{\alpha_1\alpha_2}\dots \epsilon^{\bar{\alpha}_{2n-1}\bar{\alpha}_{2n}} \epsilon^{\alpha_{2n-1}\alpha_{2n}}  + \text{all other contractions}\right)  
\label{eq: S3 int}
\end{align}

\subsection{Graviton modes}

We are interested in graviton modes $k_p$, which are in the representation $(\frac{p}{2}-1,\frac{p}{2}-1)_0$ of $SU(2)_L\times SU(2)_R\times U(1)_R$ and which have scaling dimension $\Delta=k_p$. These exist for $p=2,3,\dots$ and
\begin{align}
  k_p = \left\{ \begin{aligned}
  	\,\, p, p+2, p+4,\dots		\qquad & p=2,3			\\
  	\,\, p-2,p,p+2,\dots 		\qquad & p>3 \ .
  \end{aligned} \right.
\label{eq: kp range}
\end{align}
 Note that those operators with $k_p=p-2$ are in short multiplets, while all other modes are in long multiplets. Our task is to identify precisely the scalar spherical harmonic on $\tilde{S}^5$ which gives rise to the mode $k_p$ in AdS$_5$.

First consider the case of $S^5$ \cite{Kim:1985ez,Lee:1998bxa,Arutyunov:1998hf}. The scalar spherical harmonics on $S^5$ are labelled by a single integer $k$. The rank $k$ harmonic transforms in the rank $k$ symmetric traceless representation of $SO(6)$, and is given simply by
\begin{align}
  s_{I_1I_2\dots I_k} y_{I_1}y_{I_2}\dots y_{I_k}\ ,
\end{align}
pulled back to $S^5$, where $s_{I_1\dots I_k}$ is totally symmetric and traceless. Note that the rank $k=0,1$ modes are pure gauge (that is, can be removed by a diffeomorphism), and thus physical modes start at $k=2$. This spherical harmonic then gives rise to a scalar field $s_{I_1\dots I_k}$ in AdS$_5$ of mass $m^2 =k(k-4)$, which is thus dual to an operator of scaling dimension $\Delta=k$.
 
The orbifold then projects out certain combinations of the components of $s_{I_1,\dots,I_k}$; from what remains, we need to build irreducible representations of $SU(2)_L\times SU(2)_R\times U(1)_R$. Rather than doing this in full, note that the only graviton modes which can couple to gluon modes on the branes at the fixed point locus are precisely those which do not vanish at $\rho=0$. It is simple to see that all such modes are uncharged under $U(1)_R$. So let's focus on these modes.

Then, we find straightforwardly that the scalar spherical harmonic corresponding to the operator $k_p$ is given by
\begin{align}
  k_p:\qquad s_{i_1 i_2 \dots i_{p-2}}y_{i_1}y_{i_2}\dots y_{i_{p-2}} (1-\rho^2)^{(k_p-p+2)/2} - (SO(6) \text{ traces})\ ,
  \label{eq: kp mode}
\end{align}
where here $s_{i_1\dots i_{p-2}}$ is symmetric and traceless on its $SO(4)$ indices. Note that that the rank $n$ symmetric traceless representation of $SO(4)$ is irreducible, and is nothing other that the representation $(\frac{n}{2},\frac{n}{2})$ of $SU(2)_L\times SU(2)_R$. Thus, $k_p$ does indeed transform in the representation $(\frac{p}{2}-1,\frac{p}{2}-1)_0$ as required. As a consistency check, note that indeed the above formula makes sense only for
\begin{align}
  k_p-(p-2) = 0,2,4,\dots \ ,
\end{align}
and that for $p=2,3$ the first of these options corresponds to $k_2=0$ and $k_3=1$, respectively, which as discussed above are pure gauge and so are thrown away. So we recover precisely (\ref{eq: kp range}).

To proceed, we will need to perform the removal of traces explicitly to get a closed form for the mode in (\ref{eq: kp mode}). There is a slick way to do this. It is easy to show that after removal of traces, the mode will take the form
\begin{align}
  k_p:\qquad s_{i_1 i_2 \dots i_{p-2}}y_{i_1}y_{i_2}\dots y_{i_{p-2}} (1-\rho^2)^{(k_p-p+2)/2} f_{k_p} \left(\frac{\rho^2 }{1-\rho^2}\right)\ ,
  \label{eq: kp mode explicit}
\end{align}
for a degree $(k_p-p+2)/2$ polynomial $f_{k_p}(z)$ we must determine. But the tracelessness condition is just the statement that the mode satisfies the Laplace equation $\nabla^2(\dots) = 0$ in the ambient space $\R^6$. This implies that $f_{k_p}(z)$ obeys
\begin{align}
  (k_p - p+2)(k_p+p)f_{k_p}(z) + 4\left(1-k_pz\right) f'_{k_p}(z) + 4 z (1+z)f''_{k_p}(z) = 0\ .
\end{align}
This is solved by
\begin{align}
  f_{k_p}(z) ={}_2F_{1} \left(-\frac{k_p+p}{2}, \frac{p-k_p-2}{2},1;-z\right)\ ,
\end{align}
which is indeed a polynomial of degree $(k_p-p+2)/2$. Using some hypergeometric identities, we finally arrive at
\begin{align}
  s_{(k_p)}(x,y)&=\frac{1}{\cN(k_p)}(s_{(k_p)})_{\bar{\alpha}_1\dots \bar{\alpha}_{p-2},\alpha_1\dots \alpha_{p-2}}(x) \,y^{\bar{\alpha}_1\alpha_1}\dots y^{\bar{\alpha}_{p-2} \alpha_{p-2}}  \, {}_2 F_{1}\left(\frac{p+k_p+2}{2},\frac{p-k_p-2}{2},1;\rho^2\right)\ ,
  \label{eq: kp mode explicit2}
\end{align}
where $x$ are coordinates in AdS$_5$. We have chosen to write everything in a manifestly $SU(2)_L\times SU(2)_R$ covariant way. So, $s_{\bar{\alpha}_1\dots \bar{\alpha}_{p-2},\alpha_1\dots \alpha_{p-2}}$ is totally symmetric on both its barred and unbarred indicies. $\cN(k_p)$ is a normalisation factor that is in place to ensure that $s_{\bar{\alpha}_1\dots \bar{\alpha}_{p-2},\alpha_1\dots \alpha_{p-2}}(y)$ is a canonically normalised scalar field in AdS$_5$. We will fix it shortly.
 
We are finally ready to write down the corresponding profile for the metric and 4-form. If we turn on just the $k_p$ mode, we have \cite{Kim:1985ez,Lee:1998bxa}
\begin{align}
  \delta g_{\mu\nu} 	&= \frac{2}{k_p+1}\left(\nabla_\mu \nabla_\nu + \nabla_\nu \nabla_\mu - \tfrac{2}{5}g_{\mu\nu}\square\right)s_{(k_p)} - \frac{6}{5} k_p g_{\mu\nu} s_{(k_p)}	\ ,	\nn\\
  \delta g_{ab}		&= 2k_p g_{ab} s_{(k_p)}	\ ,	\nn\\
  \delta g_{\mu a }		&= 0		\ ,		\nn\\
  \delta C			&= \frac{1}{6} \left(\varepsilon_{\mu\nu\rho\sigma\tau}(\nabla^\tau s_{(k_p)})\,dx^\mu \wedge dx^\nu \wedge dx^\rho \wedge dx^\sigma - \varepsilon_{abcde} (\nabla^e s_{(k_p)})\,dy^a \wedge dy^b \wedge dy^c \wedge dy^d\right)\ .
\end{align}
Before moving on, we fix the all-important normalisation $\cN(k_p)$. Using the integral (\ref{eq: S5 gen int SU(2)}) we compute
\begin{align}
  &\cN(k_p)^2\int_{\tilde{S}^5} d\Omega_5 \,\, ( s_{(k_p)})^2 = \frac{2\pi^3}{(p-1)\mathbf{\Delta}} \,\frac{1}{k_p+2}(s_{(k_p)})_{\bar{\alpha}_1\dots \bar{\alpha}_{p-2}, \alpha_1 \dots \alpha_{p-2}} (s_{(k_p)})^{\bar{\alpha}_1\dots \bar{\alpha}_{p-2}, \alpha_1 \dots \alpha_{p-2}}\ ,
\end{align}
where indices are contracted with $\epsilon^{\bar{\alpha} \bar{\beta}}$ and $\epsilon^{\alpha \beta}$, and we have used the identity
\begin{align}
  \int_0^1 dt\,(1-t)^{p-1}\,\left[{}_2 F_1\!\left(\frac{p+k_p+2}{2},\frac{p-k_p-2}{2};1;t\right)\right]^2 = \frac{1}{k_p+2}\ .
\end{align}
Thus, in terms of the IIB coupling $\kappa$ we have
\begin{align}
  \cN(k_p)^2 = \frac{32\pi^3}{\mathbf{\Delta}\kappa^2}\frac{k_p(k_p-1)}{(k_p+1)(p-1)} \ ,
\label{eq: N norm}
\end{align}
where the additional factors come from the structure of the IIB Lagrangian \cite{Lee:1998bxa}. With this normalisation set, we ultimately then find the kinetic AdS$_5$ action
\begin{align}
  S_\text{kinetic}^{(k_p)} &= \int_{\text{AdS}_5} d^5 x \sqrt{-g_{\text{AdS}}} \,\,\bigg(  -\frac{1}{2} \nabla_\mu(s_{(k_p)})_{\bar{\alpha}_1\dots \bar{\alpha}_{p-2},\alpha_1 \dots \alpha_{p-2}}\nabla^\mu (s_{(k_p)})^{\bar{\alpha}_1\dots \bar{\alpha}_{p-2},\alpha_1 \dots \alpha_{p-2}}		\nn\\
  &\hspace{50mm}-\frac{1}{2}k_p(k_p-4)(s_{(k_p)})_{\bar{\alpha}_1\dots \bar{\alpha}_{p-2},\alpha_1 \dots \alpha_{p-2}} (s_{(k_p)})^{\bar{\alpha}_1\dots \bar{\alpha}_{p-2},\alpha_1 \dots \alpha_{p-2}} \bigg)	\ .
\end{align}
Let us finally state the values $(\delta g_{\mu\nu},\delta g_{\hat{a}\hat{b}},\delta g_{\mu \hat{a}})$ of the graviton fluctuations when pulled back to AdS$_5\times S^3$, where as in the main text, $\hat{a},\hat{b},\dots$ are indices on $S^3$. We have
\begin{align}
  \delta g_{\mu\nu} 	&= \frac{2}{k_p+1}\left(\nabla_\mu \nabla_\nu + \nabla_\nu \nabla_\mu - \tfrac{2}{5}g_{\mu\nu}\square\right)s_{(k_p)} - \frac{6}{5} k_p g_{\mu\nu} s_{(k_p)}	\ ,	\nn\\
  \delta g_{\hat{a}\hat{b}}		&= 2k_p g_{\hat a \hat b} s_{(k_p)}	\ ,	\nn\\
  \delta g_{\mu \hat a }		&= 0		\ ,		\nn\\
  \delta C			&= \frac{1}{6} \varepsilon_{\mu\nu\rho\sigma\tau}(\nabla^\tau s_{(k_p)})\,dx^\mu \wedge dx^\nu \wedge dx^\rho \wedge dx^\sigma \ ,
\end{align}
written in terms of the pullback to AdS$_5\times S^3$ of the scalar harmonic $s_{(k_p)}$, which takes the simple form 
\begin{align}
  \qquad s_{(k_p)}(x,y)&=\frac{1}{\cN(k_p)}(s_{(k_p)})_{\bar{\alpha}_1\dots \bar{\alpha}_{p-2},\alpha_1\dots \alpha_{p-2}}(x) y^{\bar{\alpha}_1\alpha_1}\dots y^{\bar{\alpha}_{p-2} \alpha_{p-2}} \ .
\end{align}
This follows from ${}_2 F_1(a,b;c;0)=1$.

Note that we have trivially
\begin{align}
  g^{\mu\nu}\delta g_{\mu\nu} + g^{\hat a \hat b} \delta g_{\hat a \hat b} = 0 \ .
\end{align}
It will also be useful to note that on $S^3$,
\begin{align}
  \nabla_{\hat a} \nabla^{\hat a} s_{k_p} = -p(p-2)s_{k_p} \ .
\end{align}

\subsection{Gluon modes}

The gluon modes $p$ of interest are in the representation are in the representation $(\frac{p}{2}-1,\frac{p}{2})_0$ and have scaling dimension $\Delta=p$. The relevant vector spherical harmonics viewed as 1-forms in $\R^4$, are given by
\begin{align}
  V_p &=\varphi^{(p)}_{\bar{\alpha}_1\dots \bar{\alpha}_{p-2},\beta \gamma \alpha_1 \dots \alpha_{p-2}} (x)\,y^{\bar{\alpha}_1 \alpha_1} \dots y^{\bar{\alpha}_{p-2} \alpha_{p-2}} \left((\sigma_j)^{\bar{\delta}(\beta}(\sigma_k)_{\bar{\delta}}{}^{\gamma)}\right)  y_j dy_k			\nn\\
  & =\varphi^{(p)}_{\bar{\alpha}_1\dots \bar{\alpha}_{p-2},\beta \gamma \alpha_1 \dots \alpha_{p-2}} (x)\, y^{\bar{\alpha}_1 \alpha_1} \dots y^{\bar{\alpha}_{p-2} \alpha_{p-2}} (\sigma_j)^{\bar{\delta}\beta}(\sigma_k)_{\bar{\delta}}{}^{\gamma}   y_j dy_k	\ .
\end{align}
Note that in the first line, the expression in brackets is nothing but the anti-self-dual 't Hooft matrix, written in a manifestly $SU(2)_R$ covariant form. In particular, one can check that
\begin{align}
  \left((\sigma_i)^{\bar{\delta}(\beta}(\sigma_j)_{\bar{\delta}}{}^{\gamma)}\right) = -\left((\sigma_j)^{\bar{\delta}(\beta}(\sigma_i)_{\bar{\delta}}{}^{\gamma)}\right) = -\frac{1}{2}\epsilon_{ijkl} \left((\sigma_k)^{\bar{\delta}(\beta}(\sigma_l)_{\bar{\delta}}{}^{\gamma)}\right) \ .
\end{align}
This is what tells us that we have the $(\frac{p}{2}-1,\frac{p}{2})_0$ mode, with the other mode $(\frac{p}{2},\frac{p}{2}-1)_0$ being built from the self-dual 't Hooft matrix. Note also that $V_p$ is automatically tangent to $S^3$, i.e. $y_i (V_p)_i=0$. Finally, the AdS$_5$ scalar $\varphi^{(p)}$ is in the adjoint of $G_F$, and so so is $V_p$. We suppress any flavor indices here.

So, to turn on the $p^\text{th}$ gluon mode, we simply set the AdS$_5\times S^3$ gauge field $A$ to the value
\begin{align}
  A = \frac{1}{\cB(p)} V_p^* \ ,
\end{align}
where $V_p^*$ is the pullback of $V_p$ to AdS$_5\times S^3$, while $\cB(p)$ is a normalisation factor to ensure that the scalar field  $\varphi^{(p)}_{\bar{\alpha}_1\dots \bar{\alpha}_{p-2}, \alpha_1 \dots \alpha_{p}} $ is canonically-normalised in AdS$_5$, which we will determine shortly. First, we need a few useful facts about $A$. We have
\begin{align}
  \nabla_{\hat b} \nabla^{\hat b} A_{\hat a} 				&= (2-p^2) A_{\hat a}	\ ,		\nn\\
  \epsilon_{\hat a \hat b \hat c} \nabla^{\hat b} A^{\hat c} 		&= -p A_{\hat a}	\ ,			\nn\\
  \nabla_{\hat a} A^{\hat a}						&= 0 \ .
\end{align}
The only other thing we need is the expression for the inner product of pulled-back 1-forms,
\begin{align}
  g^{\hat{a}\hat{b}}(V_p^*)_{\hat a} (V_q^*)_{\hat b} &= (V_p)_i (V_q)_i					\nn\\
  &= 2 \varphi^{(p)}_{\bar{\alpha}_1 \dots \bar{\alpha}_{p-2}, \gamma_1 \gamma_2 \alpha_1 \dots \alpha_{p-2}}\varphi^{(q)}_{\bar{\beta}_1\dots \bar{\beta}_{q-2},}{}^{\gamma_1\gamma_2}{}_{\beta_1\dots \beta_{q-2}} y^{\bar{\alpha}_1\alpha_1}\dots y^{\bar{\alpha}_{p-2}\alpha_{p-2}} y^{\bar{\beta}_1\beta_1} \dots y^{\bar{\beta}_1\beta_{q-2}}	 \ .
\end{align}
Then, working to quadratic order in fields, we find the canonically-normalised kinetic AdS$_5$ action
\begin{align}
  S_{\text{kinetic}}^{(p)} &= \mu \,\text{Tr} \int_{\text{AdS}_5\times S^3}\left(-\frac{1}{2}F\wedge \star_8 F + \frac{1}{2}C_4\wedge F \wedge F\right)\bigg|_\text{quadratic}\nn\\		
  &= \text{Tr} \int_{\text{AdS}_5}d^5 x \sqrt{-g_{\text{AdS}}}\bigg( -\frac{1}{2} \nabla_\mu(\varphi_{(p)})_{\bar{\alpha}_1\dots \bar{\alpha}_{p-2},\alpha_1 \dots \alpha_{p}}\nabla^\mu (\varphi_{(p)})^{\bar{\alpha}_1\dots \bar{\alpha}_{p-2},\alpha_1 \dots \alpha_{p}}	\nn\\
  &\hspace{45mm} -\frac{1}{2}p(p-4) (\varphi_{(p)})_{\bar{\alpha}_1\dots \bar{\alpha}_{p-2},\alpha_1 \dots \alpha_{p}} (\varphi_{(p)})^{\bar{\alpha}_1\dots \bar{\alpha}_{p-2},\alpha_1 \dots \alpha_{p}} \bigg) \ .
\end{align}
where $\mu =(2\pi l_s^2)^2 T_7$ and we have fixed
\begin{align}
  \cB(p)^2 = \frac{4\pi^2}{p-1} \mu \ .
\label{eq: B norm}
\end{align}

\subsection{All gluon-gluon-graviton couplings}
	
We would finally like to compute the coefficient of the most generic couplings between two gluon modes and one graviton mode. The most general coupling allowed by $SU(2)_L\times SU(2)_R\times U(1)_R$ invariance is
\begin{align}
  pqk_r \ ,
\end{align}
where $r$ is allowed to take any of the following $(q-1)$ values,
\begin{align}
  r = |p-q|+2,|p-q|+4,\dots,p+q-2 \ .
\label{eq: r values}
\end{align}
So we plug in the explicit forms of the corresponding fields $\varphi_{(p)},\varphi_{(q)},s_{(k_r)}$ into the brane worldvolume action following (\ref{Eq: cubic expansion brane action}). The resulting expression can be manipulated using the identities
 \begin{align}
  \nabla_\mu \nabla_\nu f_1 \nabla^\mu f_2 \nabla^\nu f_3 &= \frac{1}{4}\Big((\square^2 f_1) f_2 f_3 -  f_1 (\square^2 f_2)f_3 - f_1 f_2 (\square^2 f_3) + 2 f_1 (\square f_2)(\square f_3)\Big) + \nabla_\mu \left(\dots \right)	\ ,	\nn\\
   f_1 \nabla^\mu f_2 \nabla_\mu f_3 &= \frac{1}{2}\Big((\square f_1) f_2 f_3 -  f_1 (\square f_2)f_3 - f_1 f_2 (\square f_3)\Big) + \nabla_\mu \left(\dots \right)	\ ,
\end{align}
for any scalar functions $f_1,f_2,f_3$.

We ultimately find that the cubic contribution from these interactions to the AdS$_5$ action is given by
\begin{align}
  S_\text{cubic}^{(p,q,k_r)} &= \text{Tr} \int_{\text{AdS}_5} d^5 x \sqrt{-g_{\text{AdS}}} \nn\\
  &\hspace{15mm}\times \frac{1}{1+ \delta_{p,q}}\beta_{pqk_r}(\varphi_{(p)})_{\bar{\alpha}_1\dots \bar{\alpha}_{(p+q-r-2)/2}\bar{\beta}_{1}\dots \bar{\beta}_{(p-q+r-2)/2},\gamma \delta \alpha_1\dots \alpha_{(p+q-r-2)/2}\beta_{1}\dots \beta_{(p-q+r-2)/2}} 		\nn\\
  &\hspace{35mm}\times (\varphi_{(q)})^{\bar{\alpha}_1\dots \bar{\alpha}_{(p+q-r-2)/2}}{}_{\bar{\beta}_{(p-q+r)/2}\dots \bar{\beta}_{r-2},}{}^{\gamma \delta \alpha_1\dots \alpha_{(p+q-r-2)/2}}{}_{\beta_{(p-q+r)/2}\dots \beta_{r-2}} 			\nn\\
  &\hspace{35mm}\times (s_{(k_r)})^{\bar{\beta}_1 \dots \bar{\beta}_{r-2},\beta_1\dots \beta_{r-2}} \ .
\end{align}
where we've included a symmetry factor for the case $p=q$. We find
\begin{align}
  \beta_{pqk_r} 	&= \frac{4\pi^2\mu}{\cN(k_r) \cB(p)\cB(q)}\frac{(k_r+p-q)(k_r+q-p)(k_r+p+q-2)(k_r+p+q-4)}{k_r+1}\nn\\
  				 	& \qquad \times \frac{\Gamma(p-1)\Gamma(q-1)\Gamma(r-1)}{\Gamma\!\left(\frac{p+q-r}{2}\right)\Gamma\!\left(\frac{r+p-q}{2}\right)\Gamma\!\left(\frac{r+q-p}{2}\right)\Gamma(\frac{p+q+r-2}{2})} \ .
\end{align}
Plugging in the explicit forms for $\cN$ and $\cB$, found in (\ref{eq: N norm}) and (\ref{eq: B norm}), respectively, we arrive at our final result (\ref{eq: coupling final answer}).

\section{Details of CFT data unmixing}\label{app: unmixing}

Let us provide the full details of the unmixing of CFT data, which are summarised in Section \ref{unmix}.

\subsection{Extracting leading CFT data}

We are interested in computing the OPE coefficients and anomalous dimensions of long multiplets whose superprimaries have dimension $\Delta = 2n + \cO(1/N)$, spin $\ell=0$, and are singlets under all other symmetries. We keep $n=2,3,\dots$ generic. It is well known that bulk gluon interactions mean that to determine these superprimaries, we have to take into account non-trivial mixing between the operators\footnote{The operator $:\!\phi^A_p \square^{n-p} \phi^A_p \!:$ is constructed as follows. One first takes all of the conformal primaries of the $\frac{1}{2}$-BPS multiplet generated by $\phi_p^A$, and constructs from them all possible double trace conformal primaries of dimension $\Delta=2n$ and spin $\ell=0$. These take the form of very particular sums of composites of conformal primaries and descendants \cite{Mikhailov:2002bp,Penedones:2010ue}, or equivalently as their `conglomeration' \cite{Fitzpatrick:2011dm}. Then, a particular linear combination of these conformal primaries is a superprimary; this is $:\!\phi^A_p \square^{n-p} \phi^A_p \!:$, which we normalise to have unit 2-point function. In the particular case of $n=p$, we really do just have a composite of two $\frac{1}{2}$-BPS superprimaries.}
\begin{align}
  :\!\phi_p \square^{n-p} \phi_p \!:,\quad p=2,3,\dots, n\,,
\label{eq: DT list}
\end{align}
where recall that $\phi^A_p$ is the $\frac{1}{2}$-BPS gluon superprimary of dimension $\Delta=p$, and all implicit $G_F$ and $SU(2)_L\times SU(2)_R$ indices are contracted to form a singlet.

The graviton mode $s_{2n}$, dual to superprimary $\rho_{2n}$ of dimension $\Delta=2n + \cO(1/N)$, enters in a non-zero bulk coupling of the form $s_{2n}\phi_p^A \phi_p^A$ for each $p=2,3,\dots$. This coupling is (super-)extremal for all $p=2,3,\dots,n$. The results of \cite{Castro:2024cmf} tell us then that $\rho_{2n}$ mixes non-trivially with every operator listed in (\ref{eq: DT list}). So let us allow for such mixing in our analysis; we will shortly show that the mixing coefficients parameterising this mixing are indeed non-zero, and entirely accessible from the 4-point functions we have computed.

The first step is to expand in $1/N$ the CFT data of the unmixed superprimaries $\cO_{n,a}$, where $a=1,\dots,n$ labels the degeneracy. We have scaling dimensions
\begin{align}
  \Delta_{n,a} = 2n + \frac{1}{N} \gamma^{(1)}_{n,a} + \frac{1}{N^2} \gamma^{(2)}_{n,a} + \cO\left(\frac{1}{N^3}\right)\,,
\end{align}
and superblock OPE coefficients with $\phi_p \phi_p$ given by
\begin{align}
  \lambda_{p,p,\cO_{n,a}} = \lambda^{(0)}_{p,p,\cO_{n,a}} + \frac{1}{N}\lambda^{(1)}_{p,p,\cO_{n,a}} + \frac{1}{N^2} \lambda^{(2)}_{p,p,\cO_{n,a}} + \cO\left(\frac{1}{N^3}\right)\,.
\end{align}
Our aim is then to determine the leading data $\gamma^{(1)}_{n,a}$ and $\lambda^{(0)}_{p,p,\cO_{n,a}}$ for $p=2,\dots n$, for which we need to study $\langle ppqq \rangle $ for all $p,q=2,\dots,n$. This correlator takes the form
\es{ppqq}{
&\langle \phi_p^A(y_1,x_1) \phi_p^B(y_2,x_2) \phi_q^C(y_3,\bar y_3,x_3) \phi_q^D(y_4,\bar y_4,x_4) \rangle \nn\\
&\quad = \frac{\langle y_1,y_2\rangle^p  \langle y_3,y_4\rangle^q \langle \bar{y}_1,\bar{y}_2\rangle^{p-2} \langle \bar y_3,\bar y_4\rangle^{q-2}}{x_{12}^{2p}x_{34}^{2q}}\sum_{\bf r\in\mathfrak{g} \otimes \mathfrak{g}}G^{p,q}_{\bf r}(U,V;w)P_{\bf r}^{ABCD}\,.
}
The reduced correlator $\cG^{p,q}_{\bf r}$ is then defined by
\es{ppqq reduced}{
G^{p,q}_{\bf r}(U,V;w)=\frac{z(w-\bar z)f_{r}(\bar z)-\bar z(w-z)f_{r}(z)}{w(z-\bar z)}+\left ( 1-\frac{z}{w} \right ) \left ( 1-\frac{\bar z}{w} \right )\mathcal{G}^{p,q}_{\bf r}(U,V)\,.
}
Expanding the singlet sector reduced correlator at small $U$, we have then
\begin{align}
  &\cG^{p,q}_\mathbf{1}(U,V) \nn\\
  &= \sum_{n=2,3,\dots }U^{-1} \Bigg[\left\langle \lambda^{(0)}_{p,p,\cO_n}\lambda^{(0)}_{q,q,\cO_n}  \right\rangle \nn\\
  &\hspace{20mm} + \frac{1}{N}\left(\left\langle \lambda^{(0)}_{p,p,\cO_n}\lambda^{(1)}_{q,q,\cO_n} + \lambda^{(1)}_{p,p,\cO_n}\lambda^{(0)}_{q,q,\cO_n}  \right\rangle + \frac{1}{2}\left\langle \lambda^{(0)}_{p,p,\cO_n}\gamma_n^{(1)}\lambda^{(0)}_{q,q,\cO_n} \right\rangle  (\log U + 2\partial_\Delta)\right) \nn\\
  &\hspace{20mm} + \frac{1}{N^2} \bigg( \left\langle \lambda^{(0)}_{p,p,\cO_n} \lambda^{(2)}_{q,q,\cO_n} + \lambda^{(2)}_{p,p,\cO_n}\lambda^{(0)}_{q,q,\cO_n}  + \lambda^{(1)}_{p,p,\cO_n}\lambda^{(1)}_{q,q,\cO_n} \right\rangle		\nn\\
  &\hspace{30mm} + \frac{1}{2}\left\langle \lambda^{(0)}_{p,p,\cO_n}\gamma_n^{(1)}\lambda^{(1)}_{q,q,\cO_n} + \lambda^{(1)}_{p,p,\cO_n}\gamma_n^{(1)}\lambda^{(0)}_{q,q,\cO_n} + \lambda^{(0)}_{p,p,\cO_n}\gamma_n^{(2)}\lambda^{(0)}_{q,q,\cO_n}  \right\rangle(\log U + 2\partial_\Delta)			\nn\\
  &\hspace{30mm} + \frac{1}{8} \left\langle \lambda^{(0)}_{p,p,\cO_n}(\gamma_n^{(1)})^2\lambda^{(0)}_{q,q,\cO_n} \right\rangle	(\log^2 U + 4\log U \partial_\Delta + 4 \partial_\Delta^2) \bigg)			\nn\\
   &\hspace{20mm} +  \cO\!\left(\frac{1}{N^3}\right)\Bigg]g_{\Delta+2,0}(U,V) \bigg|_{\Delta=2n}		 		\nn\\
   &\hspace{10mm} + \dots  \,,
   \label{eq: ppqq U expn}
\end{align}
where the ellipses contain contributions from short multiplets, as well as all long multiplets with spin. Here, $\partial^\text{no log}_\Delta$ means that we take the derivative, and then throw away terms involving $\log U$, as these are already accounted for in other terms.

In (\ref{eq: ppqq U expn}), the $\cO(1)$ term is computed in generalized free field theory; the $\cO(1/N)$ term comes from gluon exchange; and the $\cO(1/N^2)$ term receives contributions from both 1-loop gluon processes and graviton exchange. In this expression, the angled brackets denote averages over the $n$-fold multiplicity of data, so that for instance
\begin{align}
  \left\langle \lambda^{(0)}_{p,p,\cO_n}\gamma_n^{(1)}\lambda^{(0)}_{q,q,\cO_n}  \right\rangle = \sum_{a=1}^n \lambda^{(0)}_{p,p,\cO_{n,a}}\gamma^{(1)}_{n,a}\lambda^{(0)}_{q,q,\cO_{n,a}}  \,.
\end{align}
We are interested here in fixing the leading non-trivial data $\gamma^{(1)}_{n,a}$ and $\lambda^{(0)}_{p,p,\cO_{n,a}}$ for $p=2,\dots,n$, which constitutes $n^2$ pieces of CFT data. We see that there are precisely three coefficients in (\ref{eq: ppqq U expn}) which depend only on these pieces of data. We shall shortly see that the knowledge of these coefficients for all $p,q=2,\dots,n$ places $(n^2-1)$ independent constraints on these leading data, with a final constraint coming from the structure of the $1/N$ expansion of the bulk theory, thus allowing us to fix these leading data completely. 

To proceed, we first note that there exist numbers $a_n^p,b_n^{p,q}=b_n^{q,p},c_n^{p}$ with $p,q=2,\dots,n$ such that\footnote{Note that this only determines the $a_n^p,b_n^{p,q},c_n^p$ up to certain sign flips, corresponding to flipping the signs of the rows and columns of the mixing matrix $M_n$. This of course just corresponds to the sign ambiguity in how we define the operators $\rho_{2n}$ and $\phi_p \square^{n-p}\phi_p$. As such, it is straightforward to see that the physical data---that is, the leading anomalous dimensions $\gamma^{(1)}_{n,a}$ and OPE coefficients squared $(\lambda^{(0)}_{p,p,\cO_{n,a}})^2$---are invariant under such sign flips.}
\begin{align}
  \left\langle \lambda^{(0)}_{p,p,\cO_n}\lambda^{(0)}_{q,q,\cO_n}  \right\rangle					&=  
  a_n^p a_n^q	\delta_{p,q}	\,,	\label{eq: 0 moment}\\
  \left\langle \lambda^{(0)}_{p,p,\cO_n}\gamma^{(1)}\lambda^{(0)}_{q,q,\cO_n} \right\rangle 	&= 
  a_n^p a_n^q b_n^{p,q}		\,,\label{eq: 1 moment}\\
  \left\langle \lambda^{(0)}_{p,p,\cO_n}(\gamma^{(1)})^2\lambda^{(0)}_{q,q,\cO_n} \right\rangle &= 
  a_n^p a_n^q \left(\sum_{s=2}^n b_n^{p,s}b_n^{q,s} + c_n^p c_n^q \right)\,.\label{eq: 2 moment}
\end{align}
Note that (\ref{eq: 0 moment}) follows from large $N$ factorisation, while (\ref{eq: 1 moment}) is a trivial rescaling. The final expression (\ref{eq: 2 moment}) can then be shown using the mutual orthogonality of the $\lambda^{(0)}_{p,p,\cO_n}$ as stated in (\ref{eq: 0 moment}). The first term comes from 1-loop gluon processes, and indeed matches the form derived in \cite{Alday:2021ajh,Huang:2023ppy}. The second term meanwhile comes from graviton exchange, and in particular $c_n^p\neq 0$ precisely if there is non-trivial mixing between $\rho_{2n}$ and $:\!\phi^A_p  \square^{n-p} \phi^A_p \!:$.

This reparameterisation of the data contained in $\langle ppqq \rangle $ is chosen so as to simplify the form of the $n\times n$ mixing matrix $M_n$, defined by
\begin{align}
  &\begin{pmatrix}
  	\langle \rho_{2n}(x) \rho_{2n}(y) \rangle 	& \langle \rho_{2n}(x) :\!\phi_q  \square^{n-q} \phi_q \!:\!(y) \rangle	\\
  	\langle :\!\phi_p  \square^{n-p} \phi_p \!:\!(x)\, \rho_{2n}(y)  \rangle & \langle :\!\phi_p  \square^{n-p} \phi_p \!:\!(x) :\!\phi_q  \square^{n-q} \phi_q \!:\!(y) \rangle		
  \end{pmatrix}\nn\\
  &\qquad = \frac{1}{|x-y|^{4n}} \left [\, \delta_{ab} - \frac{1}{N} (M_{n})_{ab} \log\left(|x-y|^2\right) + \cO\left(\frac{1}{N^2}\right) \right]\,,
\end{align}
where $p,q=2,\dots,n$ and $a,b=1,2,\dots,n$. In terms of the data we have gathered, we have simply\footnote{The basic fact that goes into showing this is that the 3-point function $\langle \phi_p \phi_p \, :\!\phi_q \square^{n-q}\phi_q\!:\rangle $ can only be non-vanishing at $\cO(1)$ if $p=q$, as follows from large $N$ factorisation.}
\begin{align}
  M_n = \begin{pmatrix}
  	d_n			&  c_n^2			& c_n^3			& \dots 		& c_n^n		\\
  	c_n^2		&	b_n^{2,2}	& b_n^{2,3}		& \dots 		& b_n^{2,n}	\\
  	c_n^3		& b_n^{3,2}		& b_n^{3,3}		& \dots 		& b_n^{3,n}	\\
  	\vdots		& \vdots			& \vdots			& \vdots		& \vdots		\\
  	c_n^n	 	& b_n^{n,2	}	& b_n^{n,3}		& \dots 		& b_n^{n,n}
  \end{pmatrix}\,.
\label{eq: generic mixing matrix}
\end{align}
All that remains is to fix the number $d_n$ in the corner. This is easy; we simply note that the leading correction to the 2-point function $\langle \rho_{2n}\rho_{2n} \rangle $ comes from 1-loop diagrams (both gluon and graviton), which begin at order $1/N^2$. Hence, we have
\begin{align}
  d_n=0\,.
\end{align}
Note that in terms of unmixed CFT data, this constraint reads
\begin{align}
  \left\langle \gamma^{(1)}_{n}\right\rangle = \sum_{p=2}^n \frac{\left\langle \lambda^{(0)}_{p,p,\cO_n}\gamma^{(1)}_{n}\lambda^{(0)}_{p,p,\cO_n}\right\rangle }{\left\langle (\lambda^{(0)}_{p,p,\cO_n})^2\right\rangle}\,.
\end{align}
With the mixing matrix $M_n$ in hand, we're good to go. Let $v_{n,1},v_{n,2},\dots,v_{n,n}\in \R^n$ be the unit-normalised eigenvectors of $M_n$ with eigenvalues $\mu_{n,1},\mu_{n,2}\dots,\mu_{n,n}$. Then, the leading unmixed CFT data are given by
\begin{align}
  \gamma^{(1)}_{n,a} 			&= \mu_{n,a}	\,,				\nn\\
  \lambda^{(0)}_{p,p,\cO_{n,a}}	&= a_n^p (v_{n,a})^p\,,
\label{eq: unmixed data answer}
\end{align}
where in the final expression, $(v_{n,a})^p$ denotes the $p^\text{th}$ component of the vector $v_{n,a}\in \R^n$. This constitutes the main result of this appendix.

\subsection{Some explicit data}

Let us finally provide some explicit expressions for some of the data $a_n^p, b_n^{p,q},c_n^p$.

The $(a_n^p)^2$ are just coefficients of a superconformal block expansion of the generalized free field contribution to $\langle pppp\rangle $. They are given by
\begin{align}
   \left\langle \lambda^{(0)}_{p,p,\cO_n}\lambda^{(0)}_{p,p,\cO_n}  \right\rangle	=(a_n^p)^2 = \frac{2(2n-1)\Gamma(n)^4}{(n^2-1)\Gamma(2n)^2} \frac{\Gamma(n+p-1)\Gamma(n+p)}{\Gamma(p)^4\Gamma(n-p+2)\Gamma(n-p+1)}\,.
\end{align}
Let us also quote a closed formula for the $\frac{1}{N}\log U$ contributions to $\langle 22qq\rangle$ coming from gluon exchange \cite{Alday:2021ajh}. To extract the contribution in the flavor singlet channel, one needs to use $f^{ADJ}f^{JBC}=2h^\vee P_\mathbf{1}^{ABCD} + \dots$ and $f^{ACJ}f^{JDB}=-2h^\vee P_\mathbf{1}^{ABCD} + \dots$, where the ellipses contain other projectors. We then find 
\begin{align}
  \left\langle \lambda^{(0)}_{2,2,\cO_n}\gamma^{(1)}\lambda^{(0)}_{q,q,\cO_n} \right\rangle =	a_n^2 a_n^q b_n^{2,q} &=  -8\frac{h^\vee}{({\bf k}/N)} \frac{(-1)^q(q-1)n(2n-1)\Gamma(n)^4 \Gamma(n+q-1)}{\Gamma(2n)^2 \Gamma(q)^2 \Gamma(n-q+1)}	\nn\\
  &=  \frac{24(1-\mathbf{\Delta})}{\mathbf{\Delta}} \frac{(-1)^q(q-1)n(2n-1)\Gamma(n)^4 \Gamma(n+q-1)}{\Gamma(2n)^2 \Gamma(q)^2 \Gamma(n-q+1)}\,,
  \label{eq: 22qq log U}
\end{align}
where in reaching the last line we have used the explicit form (\ref{eq: flavor central charge}) for ${\bf k}$ as well as the relation $h^\vee = 6(\mathbf{\Delta}-1)$.  
 
 For the remaining data $b_n^{p,q},c_n^p$, we specialise to the first few twists\footnote{As a consistency check, we have checked that if we turn off the gluon-graviton mixing coefficients $c_n^p$, then the mixing matrices derived here for twists $4$ and $6$ recover the unmixed anomalous dimensions appearing in \cite{Drummond:2022dxd} up to a factor of $(2h^\vee/\mathbf{\Delta})$. The factor of $2h^\vee$ comes from the explicit expansion into flavor irreps that was not performed explicitly there, while the factor of $\mathbf{\Delta}$ arises from the modified flavor central charge (\ref{eq: flavor central charge}) in the presence of the F-theory singularity. } .  
 
 \paragraph{Twist 4}
 
 For unmixing at twist $4$, i.e. $n=2$, we need
 \begin{align}
  a_2^2 = \sqrt{\frac{2}{3}}\,.
\end{align}
Then, from gluon exchange in $\langle 2222\rangle $ we find as a special case of (\ref{eq: 22qq log U}),
\begin{align}
   \left\langle \lambda^{(0)}_{2,2,\cO_2}\gamma^{(1)}\lambda^{(0)}_{2,2,\cO_2} \right\rangle 	&= 
  (a_2^2)^2  b_2^{2,2} = 8\frac{(1-\mathbf{\Delta})}{\mathbf{\Delta}}\,.
\end{align}
Meanwhile, from graviton exchange in $\langle 2222\rangle $ we find
\begin{align}
  \left\langle \lambda^{(0)}_{2,2,\cO_2}(\gamma^{(1)})^2\lambda^{(0)}_{2,2,\cO_2} \right\rangle_\text{grav} &= 
  (a_2^2)^2   (c_2^2)^2 = \frac{16}{5} \frac{\dim(G_F)}{\mathbf{\Delta}} = \frac{32}{5}\frac{(6-5\mathbf{\Delta})(5-6\mathbf{\Delta})}{\mathbf{\Delta}^2}\,,
\end{align}
where in reaching the final expression we have expressed $\dim(G_F)$ in terms of $\mathbf{\Delta}$. 

From these results, we straightforwardly determine $b_2^{2,2}$ and $c_2^2$, finding the mixing matrix (\ref{eq: twist 4 mixing matrix}). The resulting leading CFT data is presented in (\ref{eq: twist 4 anom dims}) and (\ref{eq: twist 4 OPE coeff}).

\paragraph{Twist 6}
For unmixing at twist $6$, i.e. $n=3$, we need
\begin{align}
  a_3^2 = \sqrt{\frac{1}{10}},\qquad a_3^3 = \sqrt{\frac{1}{4}}\,.
\end{align}
Then from gluon exchange in $\langle 2222\rangle$ and $\langle 2233\rangle $, respectively, we again use (\ref{eq: 22qq log U}) to write down
\begin{align}
   \left\langle \lambda^{(0)}_{2,2,\cO_3}\gamma^{(1)}\lambda^{(0)}_{2,2,\cO_3} \right\rangle 	&= 
  (a_3^2)^2  b_3^{2,2} = \frac{12}{5} \frac{(1-\mathbf{\Delta})}{\mathbf{\Delta}}	\,,		\nn\\
  \left\langle \lambda^{(0)}_{2,2,\cO_3}\gamma^{(1)}\lambda^{(0)}_{3,3,\cO_3} \right\rangle 	&= 
  a_3^2 a_3^3  b_3^{2,3} =	-\frac{24}{5} \frac{(1-\mathbf{\Delta})}{\mathbf{\Delta}}	\,.		
\end{align}
We also need the analogous term from $\langle 3333 \rangle$, which was found in \cite{Huang:2023ppy} to be
\begin{align}
    \left\langle \lambda^{(0)}_{3,3,\cO_3}\gamma^{(1)}\lambda^{(0)}_{3,3,\cO_3} \right\rangle 	&= 
  (a_3^3)^2  b_3^{3,3} = \frac{66}{5} \frac{(1-\mathbf{\Delta})}{\mathbf{\Delta}}\,.
\end{align}
Finally, from graviton exchange in $\langle 2222 \rangle$ and $\langle 2233 \rangle$, respectively, we find
\begin{align}
  \left\langle \lambda^{(0)}_{2,2,\cO_3}(\gamma^{(1)})^2\lambda^{(0)}_{2,2,\cO_3} \right\rangle_\text{grav} &= 
  (a_3^2)^2   (c_3^2)^2 = \frac{864}{175} \frac{\dim(G_F)}{\mathbf{\Delta}} =	\frac{1728}{175}\frac{(6-5\mathbf{\Delta})(5-6\mathbf{\Delta})}{\mathbf{\Delta}^2}	\,,	\nn\\
  \left\langle \lambda^{(0)}_{2,2,\cO_3}(\gamma^{(1)})^2\lambda^{(0)}_{3,3,\cO_3} \right\rangle_\text{grav} &= 
  a_3^2 a_3^3   c_3^2 c_3^3 = - \frac{864}{35} \frac{\dim(G_F)}{\mathbf{\Delta}} =  - \frac{1728}{35}\frac{(6-5\mathbf{\Delta})(5-6\mathbf{\Delta})}{\mathbf{\Delta}^2}\,.
\end{align}
We thus find mixing matrix
\begin{align}
  M_3 = \begin{pmatrix}
  	0		& \frac{24}{\mathbf{\Delta}}\sqrt{\frac{6}{35}}\sqrt{(6-5\mathbf{\Delta})(5-6\mathbf{\Delta})}		& -\frac{48}{\mathbf{\Delta}}\sqrt{\frac{3}{7}}\sqrt{(6-5\mathbf{\Delta})(5-6\mathbf{\Delta})}		\\
  	\frac{24}{\mathbf{\Delta}}\sqrt{\frac{6}{35}}\sqrt{(6-5\mathbf{\Delta})(5-6\mathbf{\Delta})}		& 24\frac{(1-\mathbf{\Delta})}{\mathbf{\Delta}}		& -48\sqrt{\frac{2}{5}}\frac{(1-\mathbf{\Delta})}{\mathbf{\Delta}}		\\
  	-\frac{48}{\mathbf{\Delta}}\sqrt{\frac{3}{7}}\sqrt{(6-5\mathbf{\Delta})(5-6\mathbf{\Delta})}		& -48\sqrt{\frac{2}{5}}\frac{(1-\mathbf{\Delta})}{\mathbf{\Delta}}		& \frac{264}{5}\frac{(1-\mathbf{\Delta})}{\mathbf{\Delta}}
  \end{pmatrix}\,.
\label{eq: M3}
\end{align}
Then, the leading order anomalous dimensions $\gamma^{(1)}_{3,a}$ and OPE coefficients $\lambda^{(0)}_{p,p,\cO_{3,a}}$ for $a=1,2,3$ and $p=2,3$ are given in terms of the eigenvalues of eigenvectors of $M_3$, according to (\ref{eq: unmixed data answer}).

\bibliographystyle{JHEP}
\bibliography{graviton}

\end{document}